\documentclass[useAMS,usenatbib]{mn2e}

\usepackage{fontenc}
\usepackage{amsmath,amssymb,amsbsy}
\usepackage[dvips]{graphicx}
\usepackage{multirow}

\title[New Colour---$M_\star/L$ relations with TP-AGB and dust]{New 
Colour---Mass to Light Relations: the role of the Asymptotic 
Giant Branch phase and of interstellar dust}

\author[Into \& Portinari ]{Tom Into and Laura Portinari\\
Tuorla Observatory, Department of Physics and Astronomy, University 
of Turku. V\"ais\"al\"antie 20, FIN-21500 Piikki\"o, Finland}

\begin{document}

\maketitle

\begin{abstract}
Colour-$M_\star/L$ (mass--to--light) relations are a popular recipe to derive 
stellar mass in external galaxies. Stellar mass estimates often rely on near 
infrared (NIR) photometry,
considered an optimal tracer since it is little affected by dust and by
the "frosting" effect of recent star formation episodes.
However, recent literature has highlighted that theoretical estimates 
of the NIR $M_\star/L$ ratio strongly depend on the modelling of the Asymptotic 
Giant Branch (AGB) phase.

We use the latest Padova isochrones, with detailed modelling of the 
Thermally Pulsing AGB phase, to update theoretical colour-$M_\star/L$ relations 
in the optical and NIR and discuss the consequences for the estimated stellar 
masses in external galaxies.

We also discuss the effect of attenuation by interstellar dust 
on colour-$M_\star/L$ relations in the statistical case of large galaxy samples.
\end{abstract}

\begin{keywords}
Galaxies: luminosity function, mass function; galaxies: stellar content; 
infrared: galaxies; stars: AGB and post-AGB; dust, extinction
\end{keywords}

%%%%%%%%%%%%%%%%%%%%%%%%%%%%%%%%%%%%%%%%%%%%%%%%%%%%%%%%%%%%%%%%%%%%%%%%
\section{Introduction}
Various astrophysical problems require to estimate stellar mass 
from the photometry: for instance, to reconstruct the assembly 
history of galaxies through cosmic times
\citep[e.g.][2004]{Bell_apjss_149} \nocite{Bell2004}; 
to define the stellar/baryonic mass Tully--Fisher relation, 
which is more physically meaningful than 
in any specific photometric band \citep{Bell_apj_550, McGaugh2005}; to
determine the inventory of baryons in the Universe (Fukugita, 
Hogan \& Peebles 1998; Fukugita \& Peebles 2004; McGaugh et~al.\ 2010);
\nocite{Fukugita98, Fukugita2004, McGaugh2010} and to disentangle the 
contribution of luminous versus dark matter in galactic dynamics 
\citep[e.g.][and references therein]{PortSal2010}. 

The key to it is the stellar mass--to--light ratio ($M_\star/L$).
For a Simple Stellar Population (ensemble of coeval
stars with the same chemical 
composition, formed in the same burst of star formation), $M_\star/L$ 
depends on (a) the Initial Mass Function; (b) the age and (c) the metallicity.;
for a composite stellar population, (b) and (c) become, respectively, 
age distribution (star formation history) and metallicity distribution.
Population synthesis techniques and chemo-photometric models of galaxies can 
predict theoretical $M_\star/L$ ratios (including both living stars and remnants)
associated to the photometric properties; yet only in the past decade this 
possibility
has been fully appreciated in the dynamical and extra-galactic community. 
In particular, colour---$M_\star/L$ relations have become 
a popular and handy tool to estimate stellar masses in external galaxies 
(e.g.\ Kranz, Slyz \& Rix 2003; McGaugh 2004; Kassin, de Jong \& Weiner 2006;
Bakos, Trujillo \& Pohlen 2008; Treuthard, Salo \& Buta 2009; Torres-Flores
et~al.\ 2011).
\nocite{Kranz_apj_586, McGaugh2004, Kassin_apj_643, Bakos_apjl_683, 
Treuthardt2009, Torres2011}

Early on, \citet{SarTin74} and \citet{LarTin78} reported tight relations 
between colour and $M_\star/L$ for $B,V$ bands, 
used by \citet{Tinsley81} to discuss the dark matter content in galaxies 
of different Hubble type. Those papers presented the first linear equation
relating colour and log($M_\star/L$), and the insightful
remark that, adopting a different Initial Mass Function (IMF) 
``All the models of a set could be arbitrarily moved up or down
in $\log M/L_B$ [...] The {\it slopes} of the relations would not be altered''.

After the pioneering papers of Tinsley and collaborators, the next analysis 
of $M_\star/L$ ratios from galactic models was by \citet{JabAri1992}, 
who extended colour--$M_\star/L$ relations to near infrared (NIR) bands; it is of
historical interest that, in those early models, the $M_\star/L_H$ vs.\ $(B-H)$ 
relation had a {\it negative} slope, at odds with relations in optical
bands. Their models were used by \citet{Persic93} to discuss
the dark matter content of galaxies as a function of luminosity.

These early results on $M_\star/L$ from galactic models went otherwise mostly
unnoticed. The breakthrough introduction of ``colour--$M_\star/L$ relations'' 
to the wider dynamical and extra-galactic 
community, was the extensive study of \citet{Bell_apj_550}: they
showed that a variety of disc galaxy models 
(closed box models, open models with inflows or outflows, with different 
formation age, with starbursts, etc.) all resulted in the same, robust
log--linear colour--$M_\star/L$ relations (CMLR). The tightest relations involve 
optical bands, where the notorious age--metallicity degeneracy is, for 
once, an advantage, as it concurs to keeping the relation tight.
Similar CMLR were found to hold {\it within} disc galaxy models, along 
the radial colour profiles obtained in the inside--out scenario 
\citep*{Portinari_mn_347}. Semi--empirical (as opposed to purely theoretical) 
CMLR have been derived by \citet{Bell_apjss_149} from multi-band photometry 
of galaxies in the SDSS+2MASS surveys.

A key feature of CMLR, also highlighted by \citet{Bell_apj_550}, is that the 
zero--point of the relation is set by the stellar IMF, while the slope 
of the relation is robust
versus this, and other, model assumptions. The zero--point in the optical
bands ($B,V,I$) seems to be well established \citep{Flynn2006}: models of the 
local Galactic disc based on star counts yield a colour--$M_\star/L$ datapoint 
for the ``solar cylinder'' that agrees with the normalization of typical 
Solar Neighbourhood IMFs \citep{Kroupa:1998, Chabrier_apj_554, Chabrier2002}.

To estimate stellar mass, near infrared (NIR) photometry
is most often a favoured choice:
(i) NIR luminosity, less affected than optical bands by minor recent 
star formation episodes (the ``frosting'' effect), is a better tracer of the
bulk of the stellar population. (ii) The $M_\star/L$ ratio in the NIR
varies, overall, less than in the optical \citep{Bell_apj_550} --- 
though it is not constant and totally insensitive to the star 
formation history, as often assumed in the past. 
(iii) NIR luminosity is less
affected by dust extinction. (iv) Nowadays NIR studies benefit from 
excellent large databases obtained from extensive surveys (2MASS, UKIDSS, etc.)

However, the integrated NIR light of a stellar population is heavily affected 
by the contribution of its Asymptotic Giant Branch (AGB) stars
\citep{Maraston_mn_300, Maraston_mn_362, Girardi_mn_300, Mouhcine2002, 
Mouhcine2003}.
At intermediate ages (0.3--3~Gyr) the AGB phase dominates the NIR emission,
lowering the K--band $M_\star/L$ ratio by a factor of 3--5 and inducing 
a colour transition to the red, reaching $(V-K) \geq 3$. For high redshift 
galaxies in the relevant age range ($z \sim 2$, ages 0.2--2~Gyr), population 
synthesis models including the AGB phase yield about 1~mag brighter K--band
luminosities, 2~mag redder $(V-K)$ colours, and 60\% lower stellar masses and 
ages than other models \citep{Maraston2006, Tonini2009}.

It is by now well established that, in spite of its short--lived nature, 
the complex Thermally Pulsing (TP)-AGB phase has considerable impact 
on NIR colours and $M_\star/L$, and needs to be accurately modelled. 
Major advances in this respect
have been implemented in the recent release of the Padova isochrones. 
In this paper, we explore their 
consequences on theoretical CMLR.

The paper is organized as follows. In Section~2 we discuss CMLR for Simple
Stellar Populations, comparing the previous and recent release of the Padova
isochrones, differing only in the TP-AGB phase implementation. In Section~3
we derive CMLR for composite stellar populations resulting from extended
star formation histories.
In Section~4 we derive CMLR for more detailed disc galaxy models 
with internal colour and metallicity gradients \citep{Portinari_mn_347}.

Dust can also influence CMLR relations, although to first order approximation, 
it is believed to have little impact
on optical CMLR, as the dust extinction+reddening vector runs almost
parallel to the CMLR itself \citep{Bell_apj_550}. 
The past decade saw major progress in modelling the effects
of interstellar dust on the integrated light of galaxies 
\citep[e.g.][]{Silva98,Popescu_aa_362,Piovan2006}.
We therefore revisit the role of dust on CMLR by including the attenuation 
effects derived from detailed radiative transfer models
\citep{Tuffs_aa_419}. Section~5 is thus dedicated
to the CMLR of (normal) dusty galaxies. In Section 6 we outline our summary
and conclusions.
%%%%%%%%%%%%%%%%%%%%%%%%%%%%%%%%%%%%%%%%%%%%%%%%%%%%%%%%%%%%%%%%%%%%%%%%%%%%%%%
\section{Colour--$M_\star/L$ relations for Simple Stellar Populations}

In this section we discuss CMLR for Simple Stellar Populations 
\citep[SSPs;][]{Renzini_Buzzoni_1986}. We analyze the role of the TP-AGB phase 
by comparing
SSPs based on the ``old'' \citep{Girardi_aas_141,Girardi_aa_391} and ``new''
(Marigo et~al.\ 2008; Girardi et~al.\ 2010)
\nocite{Marigo_aa_482, Girardi_apj_724} 
isochrone dataset of the Padova 
group\footnote{http://stev.oapd.inaf.it/cgi-bin/cmd}. The ``old'' isochrones
--- basically the same set used for the CMLR of 
\citet{Portinari_mn_347} --- included the simplified TP-AGB prescriptions 
of \citet{Girardi_mn_300}. These are now superseded by detailed, calibrated 
evolutionary models of the TP-AGB phase
\citep{Marigo_aa_469} that have been implemented in the ``new'' isochrone set.
Notice that the ``old'' and the ``new'' datasets differ only in the 
treatment of the TP-AGB phase, so that the onset of the AGB in the 
SSPs is still at $t \sim 10^8$~yrs, corresponding to a turnoff mass of about 
5~M$_\odot$. The Marigo TP-AGB models follow the 
detailed evolution of stars through pulse cycles, core mass and luminosity
growth, III dredge--up, hot--bottom burning nucleosynthesis and overluminosity,
conversion from M--type to C--type star, mass loss and final superwind phase. 
In their latest version \citep{Marigo_aa_469}, 
the models also consistently follow the variations of envelope opacities 
with surface chemical composition, and of pulsation mode; and adjust the mass 
loss rate accordingly. The main difference with respect to the ``old'' isochrone
set is the transition to C stars due to the III dredge--up, and their extended
red tail in the NIR HR diagram.
The models are calibrated to reproduce the luminosity function of thousands
of carbon stars in the Large and Small Magellanic Clouds (LMC and SMC), 
as well as the M-- and C--type star 
counts/lifetimes in Magellanic Cloud clusters; and successfully compare to 
a variety of observables (periods and mass loss rates; 
period--luminosity relations; initial--final mass relation). These models aim 
at a comprehensive coverage of the complex TP-AGB 
evolution; in contrast to other models, optimized for population synthesis 
purposes, that simply calibrate the luminosity contribution of M-- and C--type 
AGB stars as a function of SSP age
\citep{Maraston_mn_300, Maraston_mn_362}.

We retrieved isochrones with ages from $\log (t/\mathrm{yr})$ = 6.0 to 10.1,
in Johnson-Cousins \emph{UBVRIJHK}, Two Micron All Sky Survey (2MASS) 
\emph{JHK$_s$} and Sloan Digital Sky Survey (SDSS) \emph{ugriz} filters. 
Integrated SSP luminosities were computed populating the isochrones with
the \citet{Kroupa:1998} IMF, that is suitable to model the chemical
evolution of the Milky Way and provides the correct zero--point 
for optical CMLR \citep{Boissier99, Flynn2006}. 

The grid of isochrones and SSPs covers seven metallicities, with 
corresponding colour coding in the figures: $Z$=0.0001 (cyan), 0.0004 
(yellow), 0.001 (orange), 0.004 (blue), 0.008 (green), 0.019$=Z_{\odot}$ (red) 
and 0.03 (black). Our SSPs are based on the \citet{Marigo_aa_482} isochrones 
for $Z \geq 0.004$, where the AGB phase is accurately 
calibrated on observations in the Milky Way and Magellanic Clouds; 
for lower metallicities ($Z \leq 0.001$) we rely on the Girardi et~al.\ (2010)
release, calibrated on the resolved AGB population of old, metal poor dwarf 
galaxies and on the white dwarf masses of globular clusters. 
These very low metallicities are however of minor interest
for CMLR relevant to the general galaxy population: as noted by 
\citet{Bell_apj_550}, the chemical enrichment caused by even modest amounts 
of star formation raises the galaxy metallicity rapidly to at least $Z$=0.002 
even in a closed box case; and since the ``G dwarf problem'' appears to be 
ubiquitous both in disc galaxies like the Milky Way and in elliptical galaxies 
\citep{Bressan94}, the low--$Z$ tail of the metallicity distribution function 
is expected to be always  little populated. 

We express $M_\star/L$ ratios in solar units using the solar magnitudes
listed in Table~\ref{tab:solmag} obtained, consistently with the adopted 
isochrones, by interpolating the bolometric corrections of the Padova 
database for the corresponding solar model ($\log T_{eff}=3.762$, 
$\log g=4.432$).

%%%%%%%%%%%%%%%%%%%%%% Table 1
\begin{table}
\label{tab:solmag}
\caption{Adopted solar magnitudes in Johnson--Cousins--Glass bands, 
SDSS and 2MASS bands.}
\begin{center}
\begin{tabular}{llll}
\hline
\hline
band & $\cal M_\odot$ & band     & $\cal M_\odot$ \\
$B$  &     5.497     & $g$      & 5.144 \\
$V$  &     4.828     & $r$      & 4.676 \\
$R$  &     4.445     & $i$      & 4.569 \\
$I$  &     4.118     & $z$      & 4.553 \\
$J$  &     3.699     & $J^{2M}$  & 3.647 \\
$H$  &     3.356     & $H^{2M}$  & 3.334 \\
$K$  &     3.327     & $Ks^{2M}$ & 3.295 \\
\hline
\end{tabular}
\end{center}
\end{table}
%%%%%%%%%%%%%%%%%%%%%%

\begin{figure}
\begin{center}
\includegraphics[scale=0.25,angle=-90]{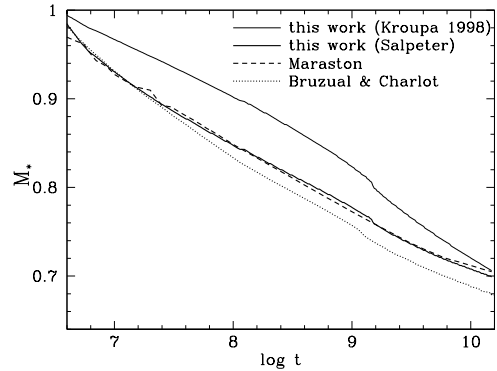}
\caption{Evolution of the SSP mass, including living stars and remnants, 
for the Kroupa (1998) IMF adopted in this work, and for the Salpeter IMF.
Also shown are the masses of the SSPs from Maraston (2005) and 
Bruzual \& Charlot (2003), with Salpeter IMF. The plot is for solar 
metallicity, but we verified that, within in each model, the impact 
of metallicity on $M_\star(t)$ is just few \%.}
\label{fig:SSPmass}
\end{center}
\end{figure}

The SSP mass $M_\star(t)$,
including both living stars and stellar remnants as a function of age,
is computed from the lifetimes and remnant masses of the Padova models
\citep*{Portinari_aa_334, Marigo2001}.
For a total initial SSP mass of 1~$M_\odot$, after a Hubble time the locked--up
fraction is typically 70\% 
for the \citet{Kroupa:1998} or the Salpeter IMF. Fig.~\ref{fig:SSPmass}
shows the evolution of the SSP mass for these two IMFs. The Salpeter case
allows a direct comparison to the SSPs by \citet{Maraston_mn_362}
and \citet{Bruzual_mn_344}; considering that the three sets of models have 
independent assumptions on stellar remnant masses and lifetimes, the agreement 
is excellent, within few~\% : any significant difference in $M_\star/L$ 
between them is entirely due to the adopted luminosity and colour evolution, 
not to the mass evolution. 
(No direct comparison is presented for the Kroupa 1998 IMF, as the public 
Kroupa SSPs by Maraston 2005 are rather based on the top--heavy IMF in Eq.~3 
of Kroupa 2001). \nocite{Kroupa_mn_322}
Notice that, since the \citet{Kroupa:1998} IMF contains fewer massive stars, 
and more long--lived stars, than the Salpeter IMF, a Kroupa SSP sheds its 
mass more slowly, although the final locked--up fraction is about the same.

%%%%%%%%%%%%%%%%%%%%%%%%%%%%%%%%%%%%%%%%%%%% Figure 1
%%%%%%%%%%%%%%%%%%%%%%%%%%%%%%%%%%%%%%%%%%%%

%%%%%%%%%%%%%%%%%%%%%%%%%%%%%%%%%%%%%%%%%%%% Figure 2
\begin{figure*}
\includegraphics[scale=0.95]{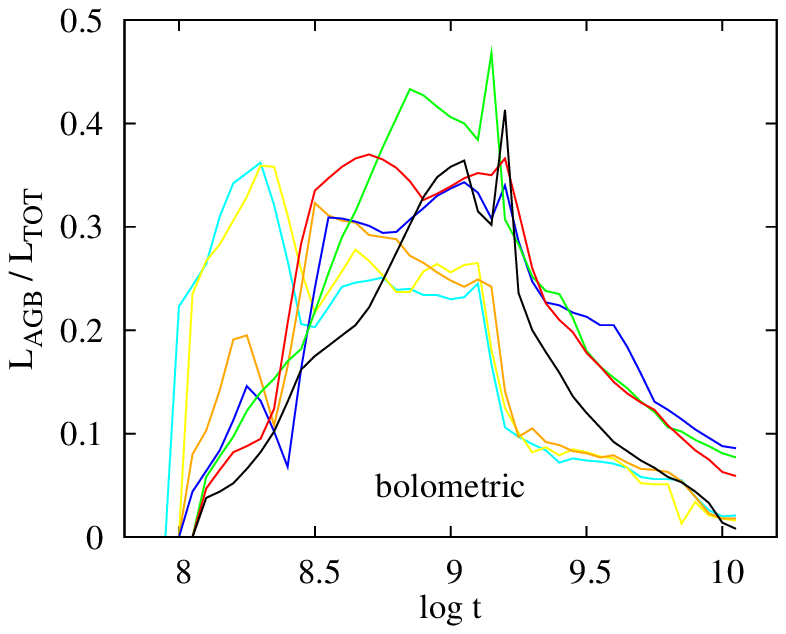}
\includegraphics[scale=0.95]{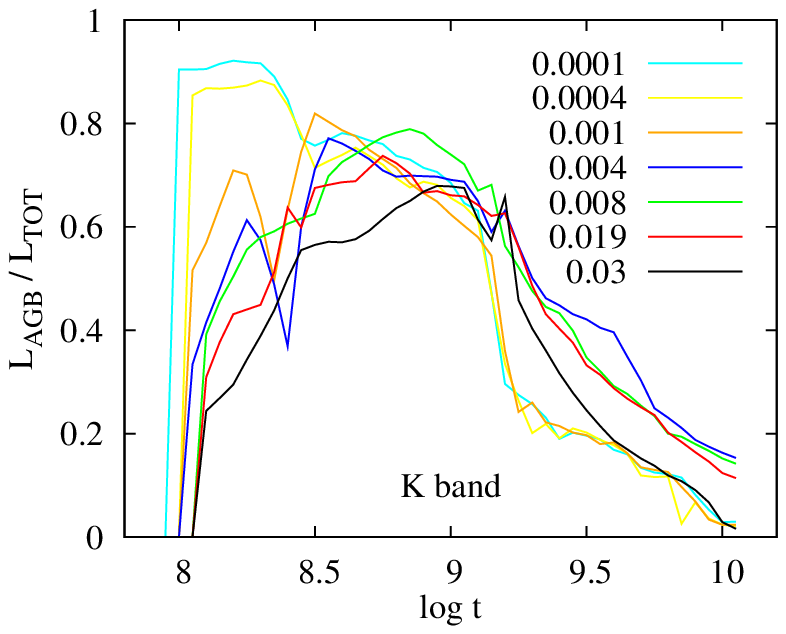}
\caption[Bolometric and $K$-band luminosity ratios for the new 
dataset]{Relative contributions of the TP-AGB phase to the total luminosity 
of a SSP using the new isochrones. The ratios are plotted 
as a function of SSP age, with metallicities listed in the legend.
\emph{Left panel}: bolometric ratios.  \emph{Right panel}: $K$-band ratios. }
\label{fig:luminosity_ratios}
\end{figure*}
%%%%%%%%%%%%%%%%%%%%%%%%%%%%%%%%%%%%%%%%%%%%

%%%%%%%%%%%%%%%%%%%%%%%%%%%%%%%%%%%%%%%%%%%%%%%%%%%%%%%%%%%%%%%%%%%%%%%%%%%%%%%
\subsection{The AGB phase contribution to the integrated light}

Updates in TP-AGB modelling are expected to affect CMLR involving NIR bands, 
where TP-AGB stars dominate, by up to 80\%, the luminosity
of SSPs of intermediate ages 0.3--3~Gyr (Figure~\ref{fig:luminosity_ratios}); 
see \citet{Maraston_mn_300, Maraston_mn_362}. This is supported 
by observations of LMC clusters \citep{Frogel_apj_352}, where AGB 
stars contribute up to 40\% of the bolometric luminosity in said age range.

\begin{figure}
\begin{center}
\includegraphics[scale=0.33]{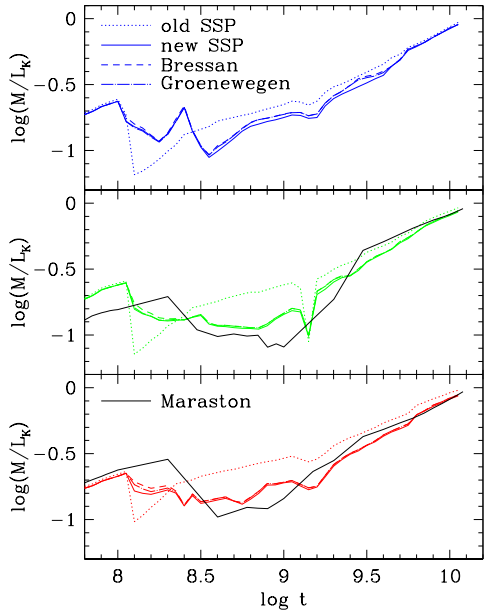}
\caption[Circumstellar dust]{$K$ band $M_\star/L$ ratio as a function of age
for the ``old'' (dotted) and ``new'' (solid) SSPs. The effect of circumstellar 
dust on the ``new'' SSPs is also shown (dashed lines). 
\emph{Top panel:} $Z$=0.004; \emph{Mid panel:} $Z$=0.008; 
\emph{bottom panel:} $Z=Z_\odot$=0.019. Notice that the luminosity peak
around $\log t=9.2$, most prominent for $Z=0.008$, is a non--spurious, 
understood feature of SSPs \citep{Girardi_mn_300}; 
it is so short-lived, however, that it is smoothed away when considering 
composite stellar populations with finite age range. For solar and LMC 
metallicity (bottom panels) we also overplot the corresponding SSPs from 
Maraston (2005)
rescaled, for the sake of this comparison, from Salpeter to Kroupa (1998) IMF
by accountig for the approriate evolutionary flux (Maraston 1998).}
\label{fig:sspdust}
\end{center}
\end{figure}

In Figure \ref{fig:sspdust}, we compare the $M_\star/L$ evolution
of the old and new SSPs in $K$ band. The onset
of the AGB phase, soon after 100~Myr, is smoother in the new models: 
the ``AGB phase transition'' is not as sharp, 
due to revised mass--loss prescriptions that reduce the lifetimes 
of the most massive TP-AGB stars \citep{Marigo_aa_482}. Later on though, 
the new models remain significantly brighter in the NIR (up to 0.5~mag), 
predicting ``lighter'' $M_\star/L$ ratios for most of the SSP lifetime. In
bluer bands, the difference between old and new models is reduced, becoming
negligible in $UBV$.

\begin{figure*}
\includegraphics[scale=1]{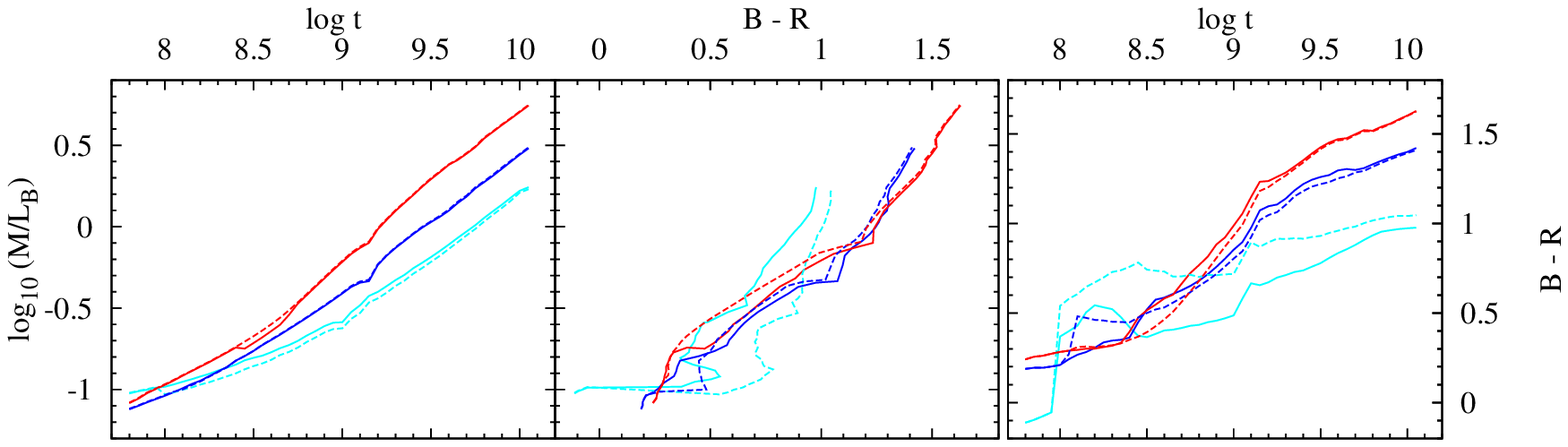}
\includegraphics[scale=1]{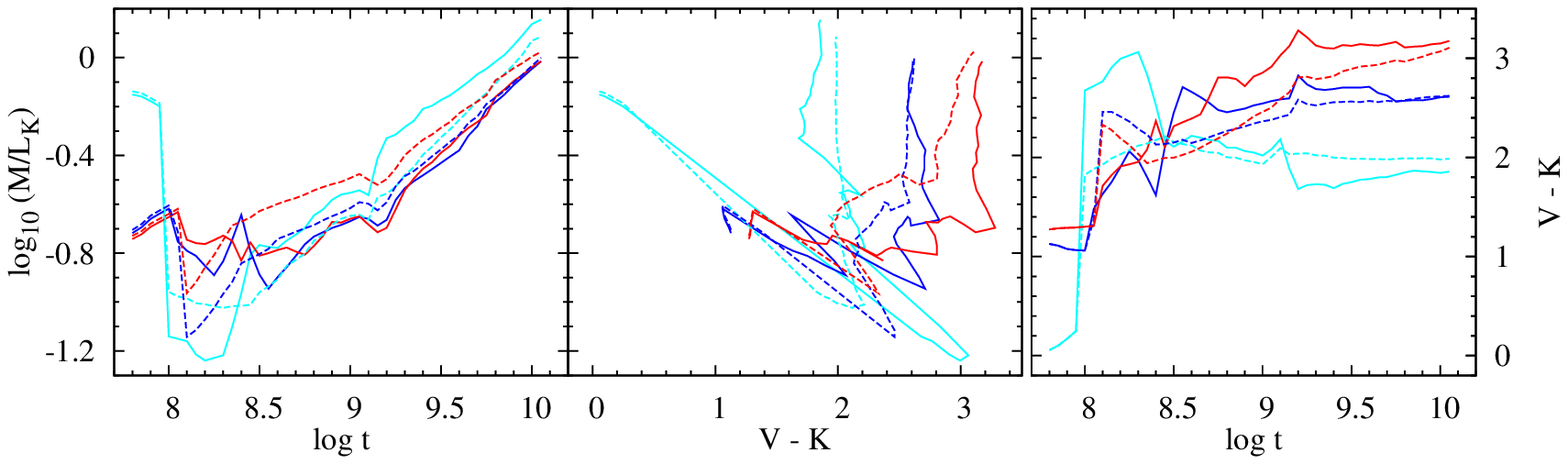}
\caption{Comparison of SSP evolution and CMLR between the old (dashed) 
and the new (solid) datasets. Three metallicities are shown: $Z$=0.0001 
(cyan), 0.004 (blue) and 0.019 (red). }
\label{fig:ssp_comparison}
\end{figure*}

As a minor detail, notice that the new $Z=0.004$ SSP presents a spike 
of high $M_\star/L_K$ (and a corresponding blue spike in $(V-K)$, see 
Fig.~\ref{fig:ssp_comparison}) around $\log t=8.4$.
As the NIR luminosity is very sensitive to the contribution of carbon stars, 
this feature can be ascribed to the complex dependence of the carbon star phase
as a function of stellar mass and age for this particular metallicity 
\citep[see Fig.~20 in][]{Marigo_aa_469}.
The spike is however smoothed away in the case of composite stellar populations.

For solar and LMC metallicity we also compare to the models of Maraston 
(2005) --- her $Z=0.02$ and~0.01 SSPs respectively: the $M_\star/L_K$ evolution
is qualitatively similar to that of the new Padova models,
although the onset of the AGB contribution occurs at $t > 200$~Myr. 
While this difference is relevant for star clusters within 
that specific age range, it is less crucial for the integrated light
and CMLR of galaxies with extended star formation histories, where the two 
models globally agree (see Fig.~5).
A detailed comparison to the Maraston models is beyond the scopes of this 
paper; some comparison can be found in \citet{Marigo_aa_482} and 
\citet{conroy2009, conroy2010} --- though notice that the latter authors adopt 
different spectral libraries, and tailor their own version of the default 
Padova models considered here.

%%%%%%%%%%%%%%%%%%%%%%%%%%%%%%%%%%%%%%%%%%%% Figure 3
%%%%%%%%%%%%%%%%%%%%%%%%%%%%%%%%%%%%%%%%%%%%

%%%%%%%%%%%%%%%%%%%%%%%%%%%%%%%%%%%%%%%%%%%%%%%%%%%%%%%%%%%%%%%%%%%%%%%%%%%%%%%
\subsection{Circumstellar dust} \label{sec:circumstellar_dust}

The new Padova isochrones can also include circumstellar dust around 
AGB stars, following the recipes of \citet{Bressan_aa_332} or 
\citet{Groenewegen_aa_448}. The circumstellar envelope reprocesses
a fraction of the stellar UV/optical light to the mid and far infrared. 
The process is crucial to interpret the MIR HR diagram of the Magellanic 
Clouds 
\citep[Marigo et~al. 2008, 2010; see also][]{Boyer2009, Kelson2010, Barmby2012} 
\nocite{Marigo2010} but NIR light is much less affected.

Figure~\ref{fig:sspdust} also compares the NIR $M_\star/L$ of new SSPs with 
and without circumstellar dust; the difference is tiny compared to that 
between the old and the new models.
Dust effects on NIR light are highlighted in two--colour 
plots such as $(J-H)$ vs.\ $(H-K)$ \citep{Bressan_aa_332, Piovan2003}, but we 
verified that, for the sake of CMLR, even for these colours the impact of dust
is smaller than the difference between the old and the new SSPs.

In bluer bands the difference between dusty and dust-free SSPs remains 
negligible, as the optical luminosity of SSPs 
is not so sensitive to the AGB phase
\citep{Maraston_mn_300, Maraston_mn_362};
so, albeit the optical emission of AGB stars does suffer from
dust reprocessing, this is of minor importance for the SSP as a whole.

In short, the impact of circumstellar dust on integrated SSP light 
is far less crucial than the improved TP-AGB modelling, 
and we shall neglect it in the remainder of this work.

%%%%%%%%%%%%%%%%%%%%%%%%%%%%%%%%%%%%%%%%%%%% Figure 4
%%%%%%%%%%%%%%%%%%%%%%%%%%%%%%%%%%%%%%%%%%%%

%%%%%%%%%%%%%%%%%%%%%%%%%%%%%%%%%%%%%%%%%%%%%%%%%%%%%%%%%%%%%%%%%%%%%%%%%%%%%%
\subsection{Time evolution of colours and mass-to-light}

Figure \ref{fig:ssp_comparison} exemplifies the origin of CMLR for SSPs, 
as the combined result of the time evolution of colours and $M_\star/L$. 
We selected for illustration three metallicities: $Z$=0.0001
is the lowest in the database; $Z$=0.004 and 0.019, 
corresponding to SMC and solar metallicity, cover the range 
significant for the bulk of stellar populations in galaxies.

The top panels show a typical optical CMLR: both $M_\star/L$ and colour smoothly
increase with SSP age and the resulting CMLR does not significantly depend 
on metallicity, at least in the range $Z=0.004-0.02$. The new implementation 
of the TP-AGB phase has very mild effects in optical bands: the AGB 
``phase transition'' is not apparent in optical colours, dominated by the 
light of turn-off Main Sequence stars and core helium--burning stars 
\citep{Renzini_Buzzoni_1986, Bressan94, Maraston_mn_300, Maraston_mn_362}. 
Only at very low metallicities, where young SSPs are much bluer, we see an
AGB colour transition.

The bottom panels in Fig.~\ref{fig:ssp_comparison} illustrate a typical 
optical--NIR CMLR, where the difference between old and new models is far more 
evident. 
The $(V-K)$ colour tends to saturate, becoming mainly a metallicity indicator, 
after $t =0.3$~Gyr. Correspondingly, the optical---NIR CMLR for SSPs breaks 
down at old ages (cf.\ the almost vertical lines in the middle panel). 
With the new SSPs, the evolution of the $K$ band $M_\star/L$ ratio is very 
similar for $0.004 \leq Z \leq 0.019$; this degeneracy with 
respect to metallicity 
was already remarked by \citet{Maraston_mn_362}.

%%%%%%%%%%%%%%%%%%%%%%%%%%%%%%%%%%%%%%%%%% Table 2
\begin{table*}
\caption[$b$ and $\tau$ values]{Grid of $b$ parameter values and corresponding 
e--folding timescales $\tau$ for our exponential models.}
\label{table_b_tau}
\begin{small}
\begin{tabular}{cccccccccccccccc}
\hline
\hline
\multirow{2}{*}{\normalsize $b$} & \vline & 0.01 & 0.05 & 0.10 & 0.15 & 0.20 & 0.30 & 0.40 & 0.50 & 0.60 & 0.70 & 0.80 & 0.90 & 1.00 \\
& \vline & 1.10 & 1.23 & 1.37 & 1.55 & 1.75 & 2.01 & 2.35 & 2.84 & 3.23 & 3.67 & 4.59 & 6.46 & 8.33 & 10.00\\
\hline
\multirow{2}{*}{\normalsize $\tau$} & \vline &1.55&2.20&2.80&3.25&3.80&4.90&6.20&8.00&10.50&15.00&23.00&50.00&$\infty$ \\
& \vline & -50.00 & -23.00 & -15.00 & -10.50 & -8.00 & -6.20 & -4.90 & -3.80 & -3.25 & -2.80 & -2.20 & -1.55 & -1.20 & -1.00 \\
\hline
\hline
\end{tabular}
\end{small}
\end{table*}
%%%%%%%%%%%%%%%%%%%%%%%%%%%%%%%%%%%%%%%%%%

%%%%%%%%%%%%%%%%%%%%%%%%%%%%%%%%%%%%%%%%%%%%%%%%%%%%%%%%%%%%%%%%%%%%%%%%%%%%%%%
\section{Colour--$M_\star/L$ relations for exponential models}
\label{sec:simple_exp_models}

Composite stellar populations --- convolutions of SSPs of different age and 
metallicity, according to a given star formation and chemical evolution history
--- are more relevant for practical applications of CMLR to real galaxies. 
In this section we shall consider exponentially declining
(or increasing) Star Formation Rates (SFR), a common recipe
to mimic the photometric properties of the Hubble sequence.

The age of our models is $T=10$~Gyr \citep[the age estimate for the Milky Way 
disc;][]{Carraro_2000}. For each metallicity, we compute 
a grid of 27 exponential models with SFR $\Psi(t) \propto e^{-t/\tau}$: 
declining SFR are modelled with e--folding timescales $\tau$ ranging from 
1.55~Gyr to $\infty$ (constant star formation rate);
increasing SFR are modelled with negative values of $\tau$ ranging from 
$-50.00$ to $-1.00$. 

These star formation histories (SFHs) can be characterized by the 
``birthrate parameter'' $b = \psi(T)/{\langle \psi \rangle}$:
the ratio between present--day and past average SFR.
This can be considered a tracer of morphological galaxy type,
with $b < 0.2$ for 
Sa--Sab discs, $b \sim 0.4$ for Sb discs and $b \sim 1$ for Sc 
discs \citep*{Kennicutt_apj_435,Sommer-Larsen_apj_596}.
Therefore, exponential models with 
$b \leq 1$ schematically represent ``normal'' spiral galaxies;
models with $b > 1$ represent blue galaxies with prominent recent star 
formation. Elliptical galaxies are 
typically well represented by old SSPs ($b \longrightarrow 0$).

The range of adopted $\tau$ and $b$ values (from 0.01 to 10.0) are tabulated 
in Table~\ref{table_b_tau}. Our grid of
exponential models is similar to that considered by \citet{Bell_apjss_149}.

%%%%%%%%%%%%%%%%%%%%%%%%%%%%%%%%%%%%%%%%%%%% Figure 5
\begin{figure*}
\begin{center}
\includegraphics[scale=0.35]{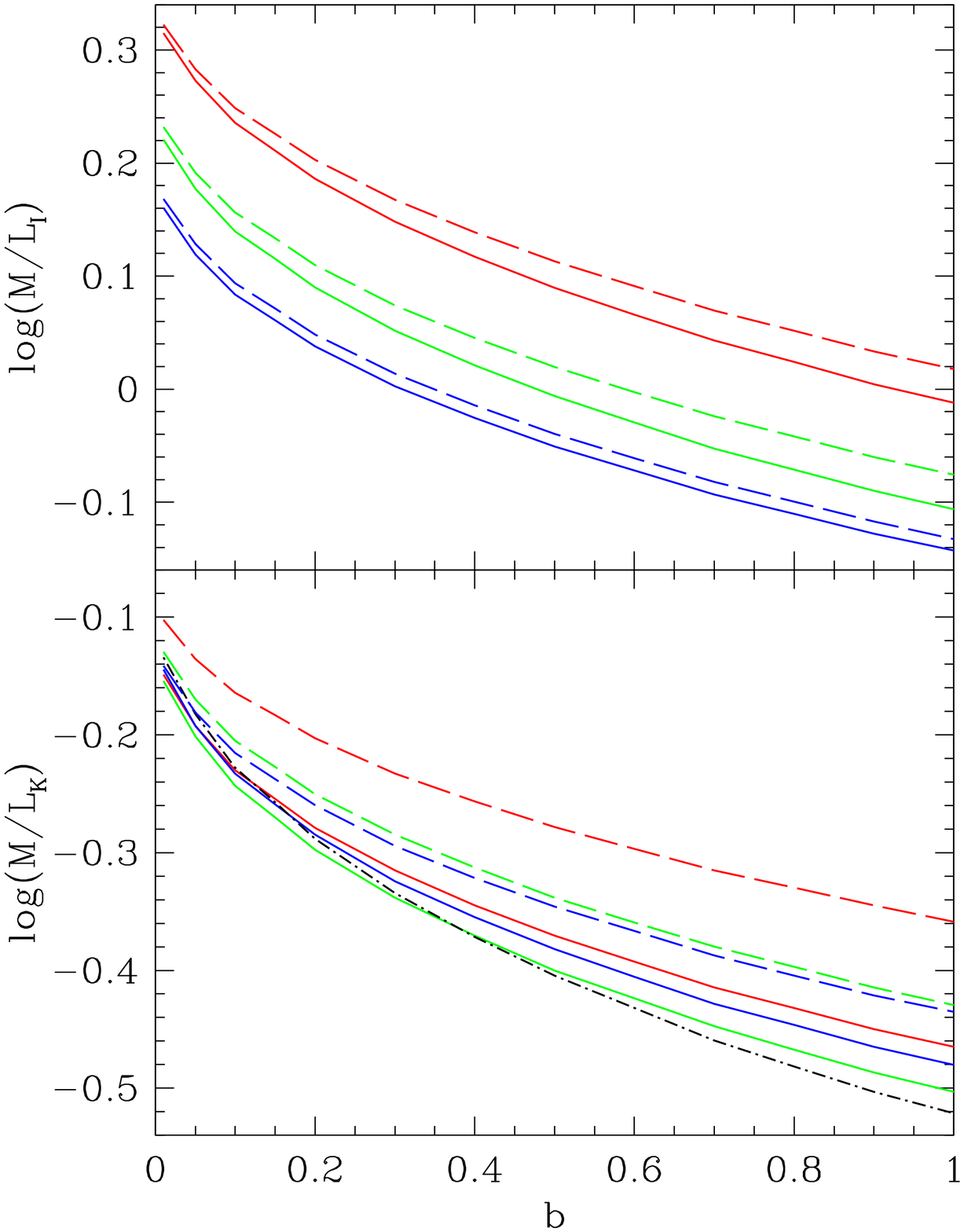}
\includegraphics[scale=0.35]{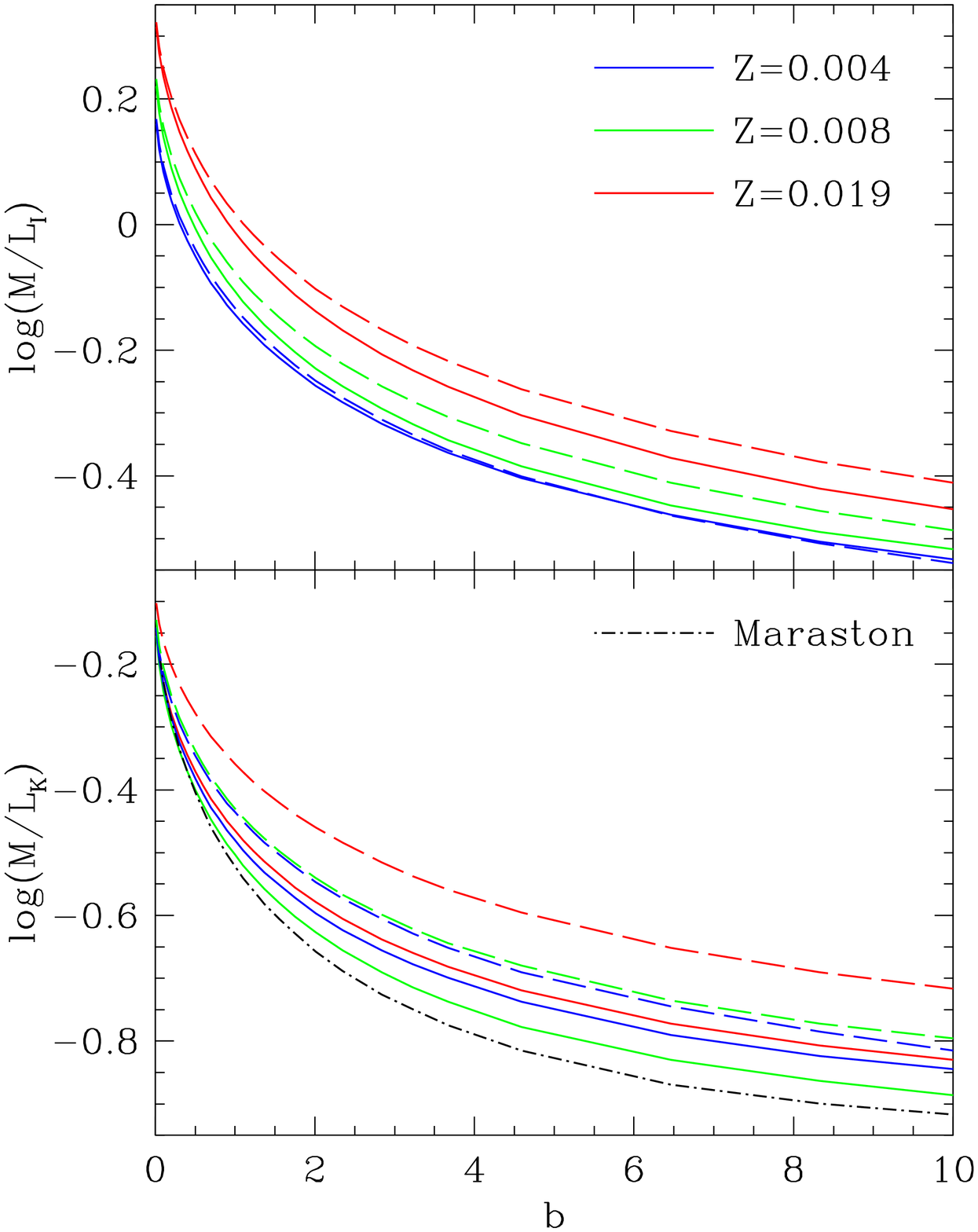}
\caption[$M/L$ versus $b$ parameter]{Mass-to-light ratio in $I$ and $K$ band
for exponential models of different metallicity, as a function of the $b$ 
parameter. Solid lines represent the new models and dashed lines the old 
models. The right panels show the full range of model SFH with $b=0.1-10$; the
left panels zoom on the more ``normal'' $b \leq 1$. The black dot--dashed line
shows the K--band $M_\star/L$ from the Maraston SSPs (Z=0.01).}
\label{fig:ML_vs_b}
\end{center}
\end{figure*}
%%%%%%%%%%%%%%%%%%%%%%%%%%%%%%%%%%%%%%%%%%%%

%%%%%%%%%%%%%%%%%%%%%%%%%%%%%%%%%%%%%%%%%%%% Figure 6
\begin{figure}
\begin{center}
\includegraphics[scale=0.9]{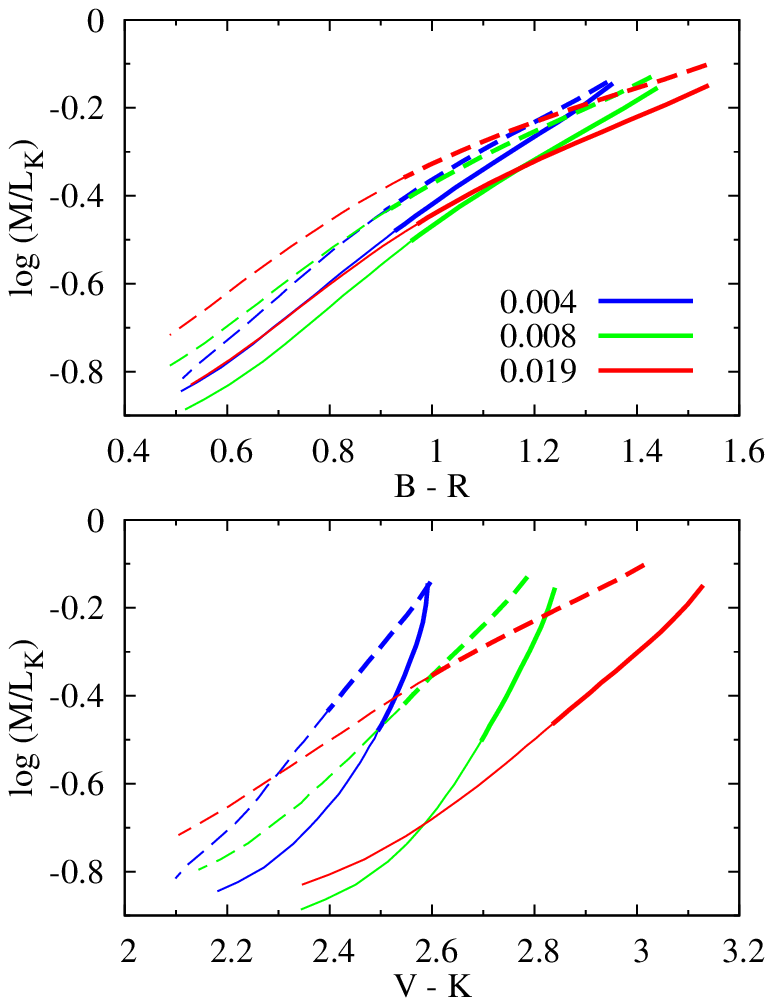}
\caption{$K$ band mass-to-light vs.\ $B-R$ (upper panel) and $V-K$ (lower 
panel). Solid/dashed lines for new/old models, as in Figure \ref{fig:ML_vs_b}.
The thick part of the curves marks the case $b \leq 1$, representative of the
SFH of ``normal'' galaxies. }
\label{fig:ML_vs_col}
\end{center}
\end{figure}
%%%%%%%%%%%%%%%%%%%%%%%%%%%%%%%%%%%%%%%%%%%%

%%%%%%%%%%%%%%%%%%%%%%%%%%%%%%%%%%%%%%%%%%%%%%%%%%%%%%%%%%%%%%%%%%%%%%%%%%%%%%%
\subsection{Effects of updated TP-AGB models}

We now compare the CMLR of exponential SFHs, derived from the old and the new 
SSPs, mostly for $Z \geq 0.004$ as the relevant metallicity range for 
integrated galaxy light.

In Figure \ref{fig:ML_vs_b} we plot the $M_\star/L$ ratio of exponential models 
versus $b$ parameter. As already shown for SSPs, the optical luminosity is only
marginally affected by the update: the new models are brighter 
just by $\leq$0.03~dex in $I$ band.
In NIR bands, the new TP-AGB implementation renders 
the models brighter up to 0.1~dex (or 25\%); the difference
is largest at solar metallicity. With the new models, the dependence of 
$M_\star/L_K$ on metallicity is largely reduced (Fig.~\ref{fig:ssp_comparison}
and Maraston 2005): NIR $M_\star/L$ ratios depend on the SFH of the system 
but not much on the underlying chemical enrichment history and metallicity 
distribution. This is convenient when deriving stellar masses from NIR 
luminosity; in comparison, for $M_\star/L_I$ the metallicity dependence 
is as strong as the SFH dependence
(at least for $b \leq 1$).

In Fig.~\ref{fig:ML_vs_col} we plot CMLR for $M_\star/L_K$ versus 
{\mbox{$(B-R)$}} and $(V-K)$ colours, a popular choice for estimating 
stellar masses \citep{Kranz_apj_586,Kassin_apj_643}.
The effect of the new SSPs on the $M_\star/L_K - (B-R)$ relation is an overall
brightening by 0.1--0.15~dex at any fixed $(B-R)$, as the latter
is quite insensitive to TP-AGB modelling (see Fig.~\ref{fig:ssp_comparison}).
This is in general the case for CMLR based on optical colours.

In optical--NIR colours instead, the new models are up to 0.3~mag redder than 
the old models. As a combined 
result of brightening and reddening, the $M_\star/L_K$ ratio {\it at a given
colour} is up to 0.3~dex, or 2 times, lighter. Factor--of--2
lighter masses were indeed derived by \citet{Maraston2006} from multi-band
photometry of galaxies at $z \sim 2$, thanks to the AGB contribution in their 
models.

Also, the $M_\star/L_K - (V-K)$ relation depends more strongly on metallicity 
for the new models, as $(V-K)$ is more of a metallicity tracer than 
a $M_\star/L$ tracer (see Fig.~\ref{fig:ssp_comparison}).
Similar comments hold for other optical--NIR colours.

\begin{figure*}
\begin{center}
\includegraphics[scale=0.9]{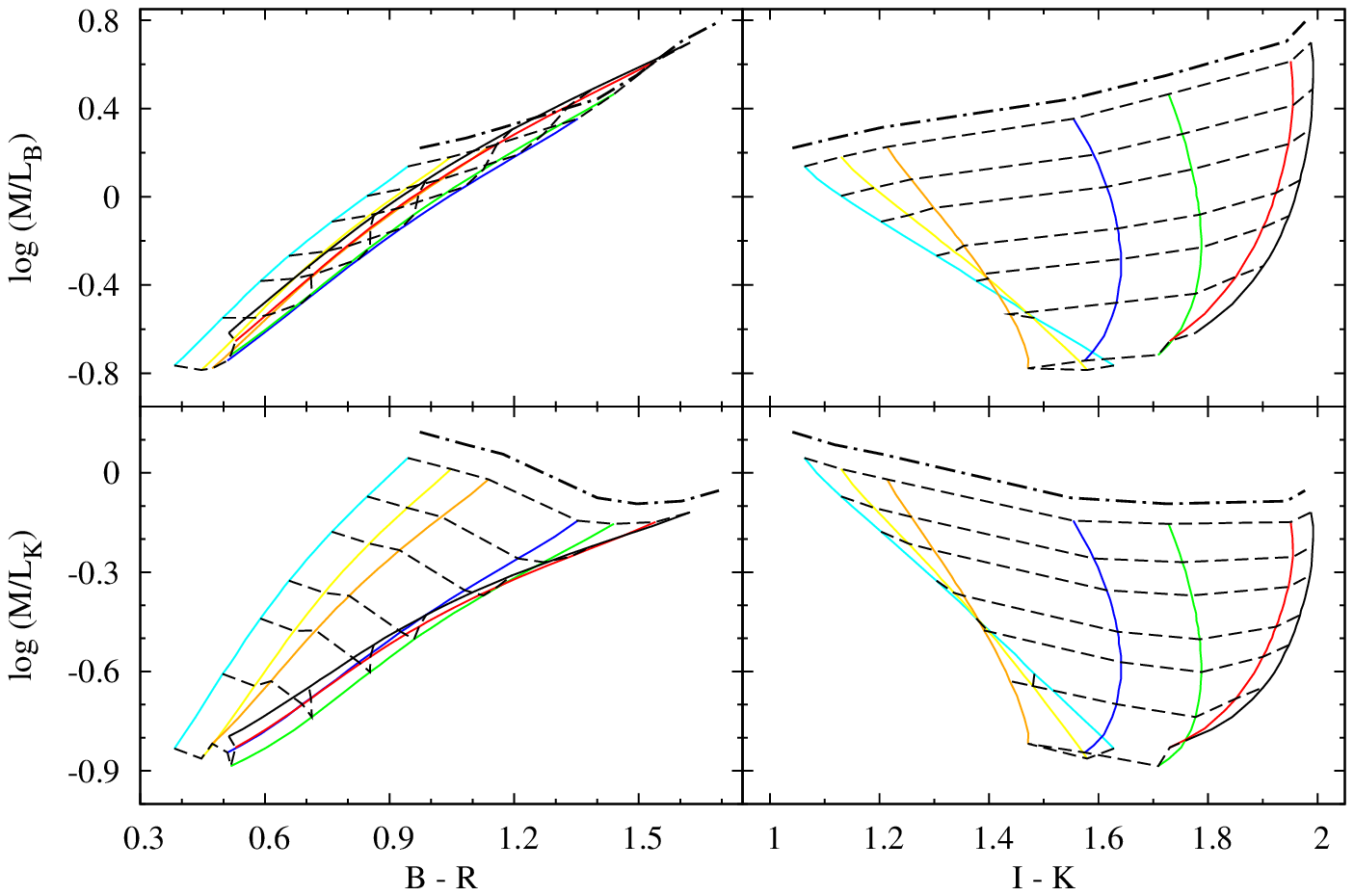}
\caption[CMLR for exponential SFH]{CMLR combining $B$ and $K$ band $M_\star/L$ 
versus $(B-R)$ and $(I-K)$ colours for the full range of $b$ parameters and
metallicities. The solid lines connect models with the same 
metallicity (colour coding as in Fig.~\protect{\ref{fig:luminosity_ratios}});
the dashed lines connect models with the same $b$ parameter: 
from top to bottom, $b=$ 0.01, 0.15, 0.40, 1.00, 1.75, 3.67, 10.00. 
The dash-dotted line represents a pure SSP of age 10 Gyr.}
\label{fig:exp_BRIK_full}
\end{center}
\end{figure*}

In Figure~\ref{fig:exp_BRIK_full} we show the new CMLR for $B$ and $K$ 
band $M_\star/L$ versus $(B-R)$ and $(I-K)$, for all the metallicities 
(coloured solid lines) and a representative sampling of the $b$--parameter 
range 0.01--10 (dashed lines).
The CMLR for a 10 Gyr old SSP is added as a dot--dashed line, smoothly extending
the CMLR of the exponential models with the lowest $b$ values.

We confirm the tight correlation between $B$ band $M_\star/L$ ratio 
and the optical $(B-R)$ colour, very little affected by SFH and metallicity 
--- other than possibly for the lowest $Z=0.0001$ case. Such a degree of 
age and metallicity degeneracy ensures robust CMLR in the optical.

CMLR remain quite robust even for $K$ band $M_\star/L$ versus $(B-R)$:
the age--metallicity degeneracy is quite tight for $Z \geq 0.004$,
breaking down only for lower metallicities that are seldom relevant in 
integrated galaxy light. 

In $(I-K)$, the age--metallicity degeneracy breaks down 
\citep{Bell_apj_550} for this is mainly a metallicity indicator, like $(V-K)$.
(For the lowest metallicities, the dependence of $(I-K)$ on $Z$ gets 
``inverted'' for $b>1$ --- i.e.\ when the system is dominated 
by young stellar populations. We ascribe this behaviour to the fact that
the extended C star phase of the more massive AGB stars causes 
a greater colour transition toward the red, the lower the metallicity; 
see Fig.~20 in Marigo \& Girardi 2007).
CMLR in $(I-K)$ are not very meaningful. The same holds 
for redder (purely NIR, e.g.\ $J-K$) colours and for $(V-K)$; while 
$(B-K)$ yields reasonably tight CMLR for $Z \geq 0.008$, the LMC metallicity.

We provide in Table~\ref{tab:fit_bmodels} log--linear CMLR for exponential 
models, for those colours that reasonably trace the $M_\star/L$ ratio,
at least in the metallicity range relevant for galaxies ($Z \geq 0.004$). 
These CMLR fit the detailed model results 
within $\pm$0.1~dex (25\% accuracy in $M_\star/L$); asterisks indicate 
less tight CMLR, accurate within $\pm$0.13~dex (35\% error).

As a rule, the fits are based on models with $0.004 \leq Z \leq 0.03$, but for 
optical--optical colours the same fits are good also down to much lower 
metallicities. In few cases, meaningful fits are limited to $Z \geq 0.008$.
The metallicity range where the fit is valid is indicated in the table.

Among optical colours, $(R-I)$ and $(r-i)$ are the least reliable $M_\star/L$
indicators; their use is further discouraged by their short baseline, as the 
error on the colour will have a significant impact on the practical estimate 
of $M_\star/L$ \citep[see also][]{Gallazzi_Bell2009}.

In Table~\ref{tab:fit_bmodels} we also provide CMLR for SDSS colours; 
they are steeper --- lighter at the blue end --- than those 
of \citet{Bell_apjss_149}. 
This is partly due to the fact that their semi--empirical relations 
are intrinsically flatter than the theoretical ones (BdJ or PST04), partly 
due to the fact that our new models result in somewhat steeper 
CMLR even for optical colours (see Section~\ref{sect:PST04}).
%The largest difference is found for the $(r-i)$ CMLR.

%%%%%%%%%%%%%%%%%%%%%%%%%%%%%%%%%%%%%%%%%% Table 3
\begin{table*}
\caption{Fitting coefficients of CMLR of the form : 
$\log (M_\star/L) = s \times {\rm colour} + z$ from exponential SFH models
with Kroupa IMF; the colour range covered by the models is indicated at 
the bottom of each sub-table.
The rightmost column indicates the metallicity range $Z \geq Z_{min}$ where 
the fitted CMLR apply (see text). Asterisks indicate less tight CMLR (accuracy
of $\pm$0.13~dex rather than $\pm$0.1~dex).}
\label{tab:fit_bmodels}
\begin{scriptsize}
\begin{tabular}{lcccl}
\hline
\hline
colour  & $\log (M_\star/L)$ &  $s$  &   $z$   & \multicolumn{1}{c}{$Z_{min}$} \\
\hline
$(B-V)$ &        $B$        & 1.866 & -1.075  & 0.0004 \\
$(B-V)$ &        $V$        & 1.466 & -0.807  & 0.0004 \\
$(B-V)$ &        $R$        & 1.277 & -0.720  & 0.0004 \\
$(B-V)$ &        $I$        & 1.147 & -0.704  & 0.0004 \\
$(B-V)$ &        $J$        & 1.047 & -0.850  & 0.004  \\
$(B-V)$ &        $H$        & 1.030 & -0.992  & 0.004  \\
$(B-V)$ &        $K$        & 1.064 & -1.066  & 0.004  \\
$(B-V)$ &      $J^{2M}$      & 1.031 & -0.835  & 0.004  \\
$(B-V)$ &      $H^{2M}$      & 1.024 & -0.991  & 0.004  \\
$(B-V)$ &      $Ks^{2M}$     & 1.055 & -1.066  & 0.004  \\
0.2--1.0 & & & & \\
\hline
$(B-R)$ &        $B$        & 1.272 & -1.287  & 0.0004 \\
$(B-R)$ &        $V$        & 1.000 & -0.975  & 0.0004 \\
$(B-R)$ &        $R$        & 0.872 & -0.866  & 0.0004 \\
$(B-R)$ &        $I$        & 0.783 & -0.836  & 0.001  \\
$(B-R)$ &        $J$        & 0.714 & -0.969  & 0.004  \\
$(B-R)$ &        $H$        & 0.701 & -1.109  & 0.004  \\
$(B-R)$ &        $K$        & 0.724 & -1.186  & 0.004  \\
$(B-R)$ &     $J^{2M}$       & 0.702 & -0.952  & 0.004  \\
$(B-R)$ &     $H^{2M}$       & 0.697 & -1.107  & 0.004  \\
$(B-R)$ &     $Ks^{2M}$      & 0.718 & -1.185  & 0.004  \\
0.5--1.6 & & & & \\
\hline
$(B-I)$ &        $B$        & 1.041 & -1.549  & 0.001 \\
$(B-I)$ &        $V$        & 0.819 & -1.182  & 0.001 \\
$(B-I)$ &        $R$        & 0.714 & -1.047  & 0.001 \\
$(B-I)$ &        $I$        & 0.641 & -0.997  & 0.004 \\
$(B-I)$ &        $J$        & 0.582 & -1.112  & 0.004 \\
$(B-I)$ &        $H$        & 0.571 & -1.249  & 0.004 \\
$(B-I)$ &        $K$        & 0.589 & -1.330  & 0.004 \\
$(B-I)$ &     $J^{2M}$       & 0.572 & -1.094  & 0.004 \\
$(B-I)$ &     $H^{2M}$       & 0.567 & -1.245  & 0.004 \\
$(B-I)$ &     $Ks^{2M}$      & 0.583 & -1.327  & 0.004 \\
0.7--2.2 & & & & \\
\hline
$(B-K)$ &        $B$        & 0.898 & -3.009  & 0.008 \\
$(B-K)$ &        $V$        & 0.710 & -2.335  & 0.008 \\
$(B-K)$ &        $R$        & 0.620 & -2.054  & 0.008 \\
$(B-K)$ &        $I$        & 0.556 & -1.901  & 0.008 \\
$(B-K)$* &       $J$        & 0.498 & -1.927  & 0.008 \\
$(B-K)$* &       $H$        & 0.487 & -2.047  & 0.008 \\
$(B-K)$* &       $K$        & 0.498 & -2.141  & 0.008 \\
2.6--4.2 & & & & \\
\hline
$(B-Ks)$ &       $B$        & 0.894 & -3.047  & 0.008 \\
$(B-Ks)$ &       $V$        & 0.707 & -2.367  & 0.008 \\
$(B-Ks)$ &       $R$        & 0.618 & -2.082  & 0.008 \\
$(B-Ks)$ &       $I$        & 0.555 & -1.928  & 0.008 \\
$(B-Ks)$* &    $J^{2M}$      & 0.489 & -1.922  & 0.008 \\
$(B-Ks)$* &    $H^{2M}$      & 0.484 & -2.069  & 0.008 \\
$(B-Ks)$* &    $Ks^{2M}$     & 0.494 & -2.166  & 0.008 \\
2.6--4.2 & & & & \\
\hline
$(V-R)$  &       $B$        & 3.963 & -1.728  & 0.004 \\
$(V-R)$  &       $V$        & 3.124 & -1.325  & 0.004 \\
$(V-R)$  &       $R$        & 2.724 & -1.172  & 0.004 \\
$(V-R)$  &       $I$        & 2.443 & -1.109  & 0.004 \\
$(V-R)$  &       $J$        & 2.220 & -1.215  & 0.004 \\
$(V-R)$  &       $H$        & 2.178 & -1.349  & 0.004 \\
$(V-R)$  &       $K$        & 2.248 & -1.434  & 0.004 \\
$(V-R)$  &    $J^{2M}$       & 2.186 & -1.195  & 0.004 \\
$(V-R)$  &    $H^{2M}$       & 2.166 & -1.346  & 0.004 \\
$(V-R)$  &    $Ks^{2M}$      & 2.228 & -1.431  & 0.004 \\
0.25--0.65 & & & & \\
\hline
 & & & & \\
 & & & & \\
 & & & & \\
\end{tabular}
\begin{tabular}{lccccc}
\hline
\hline
colour  & $\log (M_\star/L)$ &  $s$  &   $z$   & \multicolumn{1}{c}{$Z_{min}$} \\
\hline
$(V-I)$  &       $B$        & 2.312 & -2.111  & 0.004 \\
$(V-I)$  &       $V$        & 1.826 & -1.629  & 0.004 \\
$(V-I)$  &       $R$        & 1.593 & -1.438  & 0.004 \\
$(V-I)$  &       $I$        & 1.426 & -1.346  & 0.004 \\
$(V-I)$  &       $J$        & 1.285 & -1.420  & 0.004 \\
$(V-I)$* &       $H$        & 1.257 & -1.547  & 0.004 \\
$(V-I)$* &       $K$        & 1.296 & -1.637  & 0.004 \\
$(V-I)$  &    $J^{2M}$       & 1.265 & -1.397  & 0.004 \\
$(V-I)$* &    $H^{2M}$       & 1.249 & -1.541  & 0.004 \\
$(V-I)$* &    $Ks^{2M}$      & 1.282 & -1.630  & 0.004 \\
0.6--1.2 & & & & \\
\hline
$(R-I)$* &       $B$        & 5.436 & -2.594  & 0.004 \\
$(R-I)$* &       $V$        & 4.304 & -2.016  & 0.004 \\
$(R-I)$* &       $R$        & 3.756 & -1.776  & 0.004 \\
$(R-I)$* &       $I$        & 3.357 & -1.645  & 0.004 \\
$(R-I)$* &       $J$        & 2.955 & -1.674  & 0.008 \\
$(R-I)$* &       $H$        & 2.898 & -1.804  & 0.008 \\
$(R-I)$* &       $K$        & 2.965 & -1.891  & 0.008 \\
$(R-I)$* &       $J^{2M}$    & 2.934 & -1.661  & 0.008 \\
$(R-I)$* &       $H^{2M}$    & 2.907 & -1.813  & 0.008 \\
$(R-I)$* &       $Ks^{2M}$   & 2.975 & -1.906  & 0.008 \\
0.35--0.60 & & & & \\
\hline
$(g-r)$ &        $g$        & 1.774 & -0.783  & 0.0004 \\
$(g-r)$ &        $r$        & 1.373 & -0.596  & 0.0004 \\
$(g-r)$ &        $i$        & 1.227 & -0.576  & 0.001  \\
$(g-r)$ &        $z$        & 1.158 & -0.619  & 0.001  \\
$(g-r)$ &        $J^{2M}$    & 1.068 & -0.728  & 0.004  \\
$(g-r)$ &        $H^{2M}$    & 1.060 & -0.884  & 0.004  \\
$(g-r)$ &        $Ks^{2M}$   & 1.091 & -0.956  & 0.004  \\
0.1--0.85 & & & & \\
\hline
$(g-i)$ &        $g$        & 1.297 & -0.855  & 0.001  \\
$(g-i)$ &        $r$        & 1.005 & -0.652  & 0.001  \\
$(g-i)$ &        $i$        & 0.897 & -0.625  & 0.001  \\
$(g-i)$ &        $z$        & 0.845 & -0.665  & 0.004  \\
$(g-i)$ &        $J^{2M}$    & 0.779 & -0.769  & 0.004  \\
$(g-i)$ &        $H^{2M}$    & 0.772 & -0.924  & 0.004  \\
$(g-i)$ &        $Ks^{2M}$   & 0.794 & -0.997  & 0.004  \\
0.1--1.2 & & & & \\
\hline
$(g-z)$ &        $g$        & 1.152 & -0.991  & 0.004  \\
$(g-z)$ &        $r$        & 0.896 & -0.759  & 0.004  \\
$(g-z)$ &        $i$        & 0.800 & -0.721  & 0.004  \\
$(g-z)$ &        $z$        & 0.752 & -0.754  & 0.004  \\
$(g-z)$ &        $J^{2M}$    & 0.683 & -0.844  & 0.004  \\
$(g-z)$* &       $H^{2M}$    & 0.673 & -0.995  & 0.004  \\
$(g-z)$* &       $Ks^{2M}$   & 0.691 & -1.069  & 0.004  \\
0.3--1.5 & & & & \\
\hline
$(r-i)$* &        $g$        & 4.707 & -1.025  & 0.004  \\
$(r-i)$* &        $r$        & 3.650 & -0.784  & 0.004  \\
$(r-i)$* &        $i$        & 3.251 & -0.741  & 0.004  \\
$(r-i)$* &        $z$        & 3.054 & -0.773  & 0.004  \\
$(r-i)$* &        $J^{2M}$    & 2.806 & -0.867  & 0.004  \\
$(r-i)$* &        $H^{2M}$    & 2.768 & -1.018  & 0.004  \\
$(r-i)$* &        $Ks^{2M}$   & 2.843 & -1.093  & 0.004  \\
0.07--0.34 & & & & \\
\hline
$(r-z)$ &        $g$        & 3.101 & -1.323  & 0.008  \\
$(r-z)$ &        $r$        & 2.395 & -1.007  & 0.008  \\
$(r-z)$ &        $i$        & 2.128 & -0.937  & 0.008  \\
$(r-z)$ &        $z$        & 1.995 & -0.958  & 0.008  \\
$(r-z)$ &        $J^{2M}$    & 1.864 & -1.067  & 0.008  \\
$(r-z)$ &        $H^{2M}$    & 1.843 & -1.223  & 0.008  \\
$(r-z)$ &        $Ks^{2M}$   & 1.885 & -1.302  & 0.008  \\
0.2--0.62 & & & & \\
\hline
\end{tabular}
\end{scriptsize}
\end{table*}
%%%%%%%%%%%%%%%%%%%%%%%%%%%%%%%%%%%%%%%%%%

%%%%%%%%%%%%%%%%%%%%%%%%%%%%%%%%%%%%%%%%%%%% Figure 7
%%%%%%%%%%%%%%%%%%%%%%%%%%%%%%%%%%%%%%%%%%%%

%%%%%%%%%%%%%%%%%%%%%%%%%%%%%%%%%%%%%%%%%%%%%%%%%%%%%%%%%%%%%%%%%%%%%%%%%%%%%%%
\subsection{Comparison to old literature and to the no--TPAGB case}

\begin{figure}
\begin{center}
\includegraphics[scale=0.56]{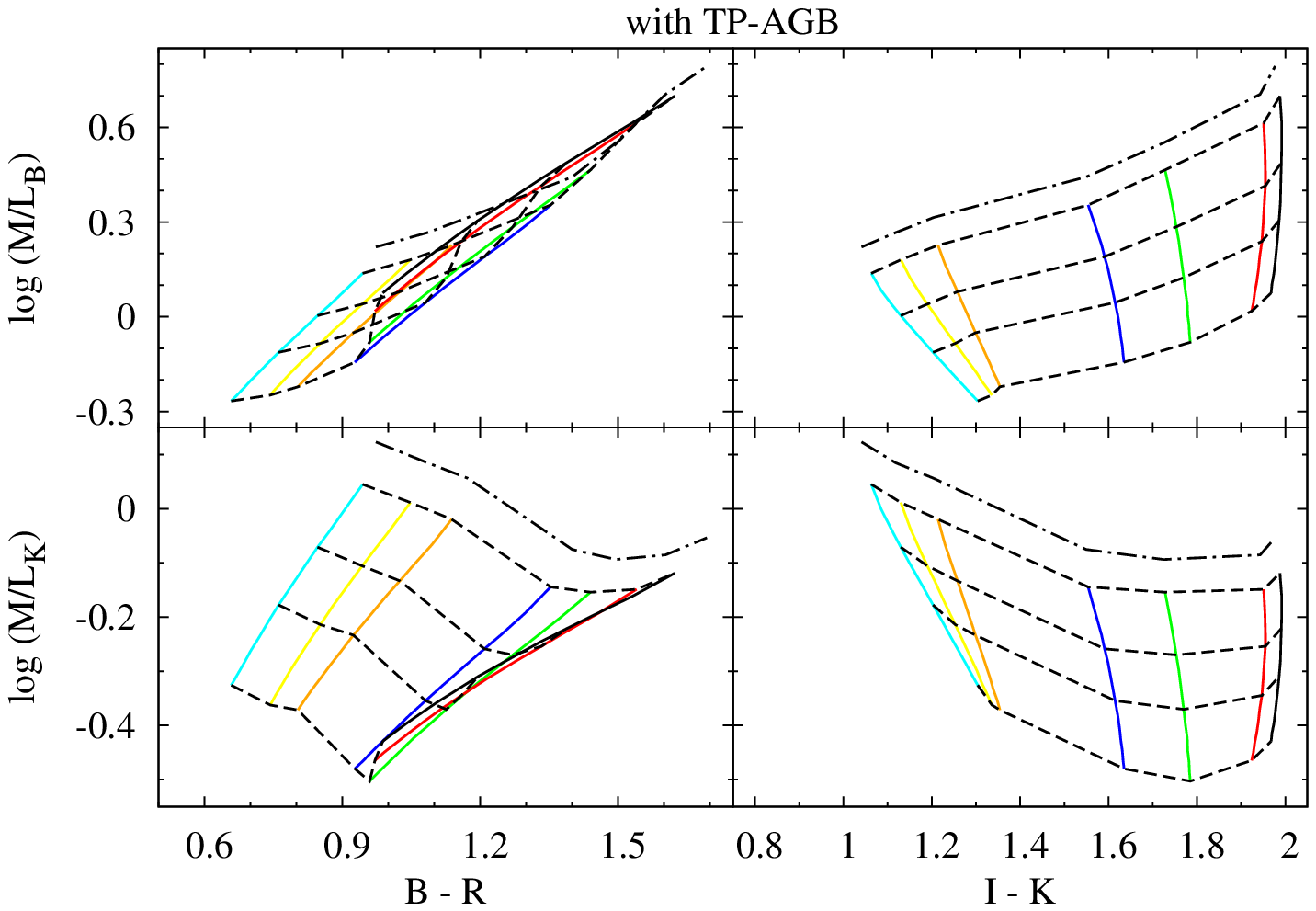}
\vspace{0.5truecm}
\includegraphics[scale=0.56]{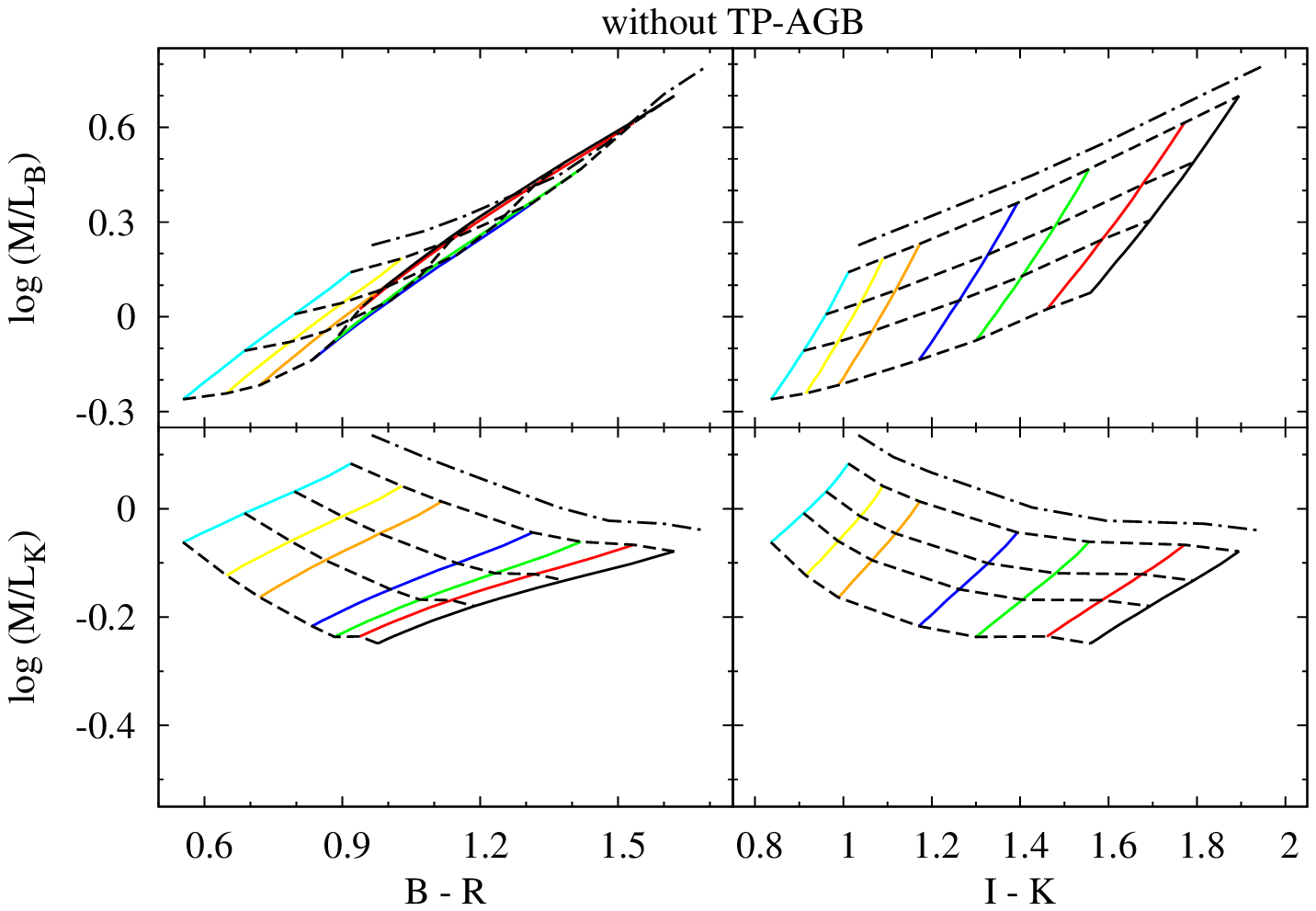}
\caption[CMLR with limited $b$ parameter values]{Same as 
Fig.~\ref{fig:exp_BRIK_full}, but limited to $b \leq 1$.
In the lower panel, the TP-AGB phase has been excluded.}
\label{fig:exp_BRIK}
\end{center}
\end{figure}

To highlight the importance of the TP-AGB phase for CMLR, we compare
to the previous theoretical relations of \citet[][hereafter BdJ]{Bell_apj_550}, 
based on an early release of the GALAXEV population synthesis package by 
\citet{Bruzual_mn_344}. Our Fig.~\ref{fig:exp_BRIK} mimics Fig.~2 of BdJ 
by limiting to exponentially decreasing SFHs ($b \leq 1$, representative
of ``normal'' galaxies) and mostly plotting the same metallicities (albeit 
their plot extends up to $Z=0.05$ and ours down to $Z=0.0001$). The two figures
are directly comparable, safe for a systematic offset in $M_\star/L$ zero-point 
of about 0.2~dex, due to the different IMF (Salpeter vs.\ Kroupa) and 
older age (12 vs.\ 10~Gyr) in their models.

The bottom panel of Fig.~\ref{fig:exp_BRIK} excludes the contribution of 
the TP-AGB phase (we computed no--TPAGB SSPs from the Padova 
isochrones, halting integration at the mass point where the Mhec flag marks 
the extension to the TP-AGB phase).
The outcome is remarkably similar to Fig.~2 of BdJ, reasserting that early 
versions of GALAXEV effectively neglected the TP-AGB contribution 
\citep{Maraston_mn_362},
although this has improved in later versions \citep{Bruzual_aspc_374}.

With respect to the no--TPAGB case, the optical CMLR becomes marginally 
less tight in the range $Z=0.004-0.03$, while $(I-K)$ becomes an even 
more neat metallicity indicator, independent of SFH.
Most interesting is the $M_\star/L_K$--$(B-R)$ CMLR: 
the ``wedge'' pattern in the no--TPAGB case, also seen in BdJ, significantly 
changes when the TP-AGB phase is included: this CMLR becomes lighter,
steeper, and tighter in the metallicity range $Z=0.004-0.03$.
The same applies to NIR $M_\star/L$ versus optical colours in general.

%%%%%%%%%%%%%%%%%%%%%%%%%%%%%%%%%%%%%%%%%%%% Figure 8
%%%%%%%%%%%%%%%%%%%%%%%%%%%%%%%%%%%%%%%%%%%%

%%%%%%%%%%%%%%%%%%%%%%%%%%%%%%%%%%%%%%%%%%%% Figure 9
\begin{figure}
\begin{center}
\includegraphics[scale=0.4]{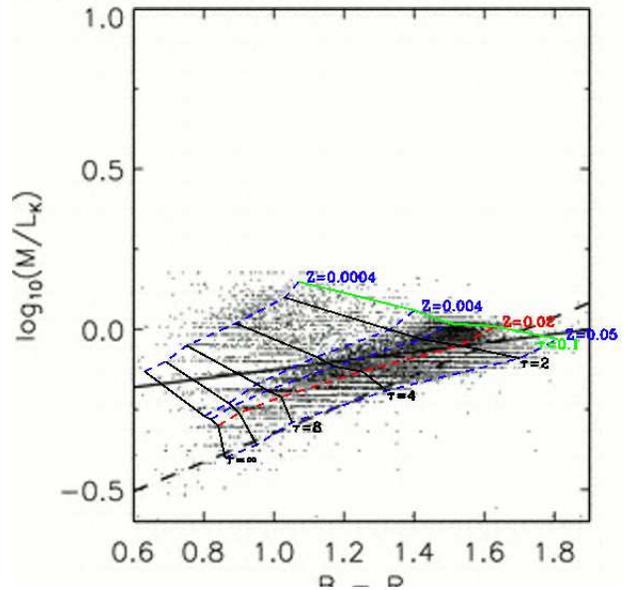}
\caption{Reproduction of Fig.~20 by \citet{Bell_apjss_149}: location of 
galaxies with multi-band SDSS+2MASS photometry in the $M/L_K$ vs.\ $(B-R)$ 
plane; the solid line is the ``semi--empirical'' CMLR derived by Bell et~al.\
the dashed line is the theoretical relation of BdJ.
Overlayed is a grid of exponential models obtained by BdJ with population
synthesis models available at the time.}
\label{fig:bell03}
\end{center}
\end{figure}
%%%%%%%%%%%%%%%%%%%%%%%%%%%%%%%%%%%%%%%%%%%%

%%%%%%%%%%%%%%%%%%%%%%%%%%%%%%%%%%%%%%%%%%%% Figure 10
\begin{figure*}
\begin{center}
\includegraphics[scale=0.62,angle=270]{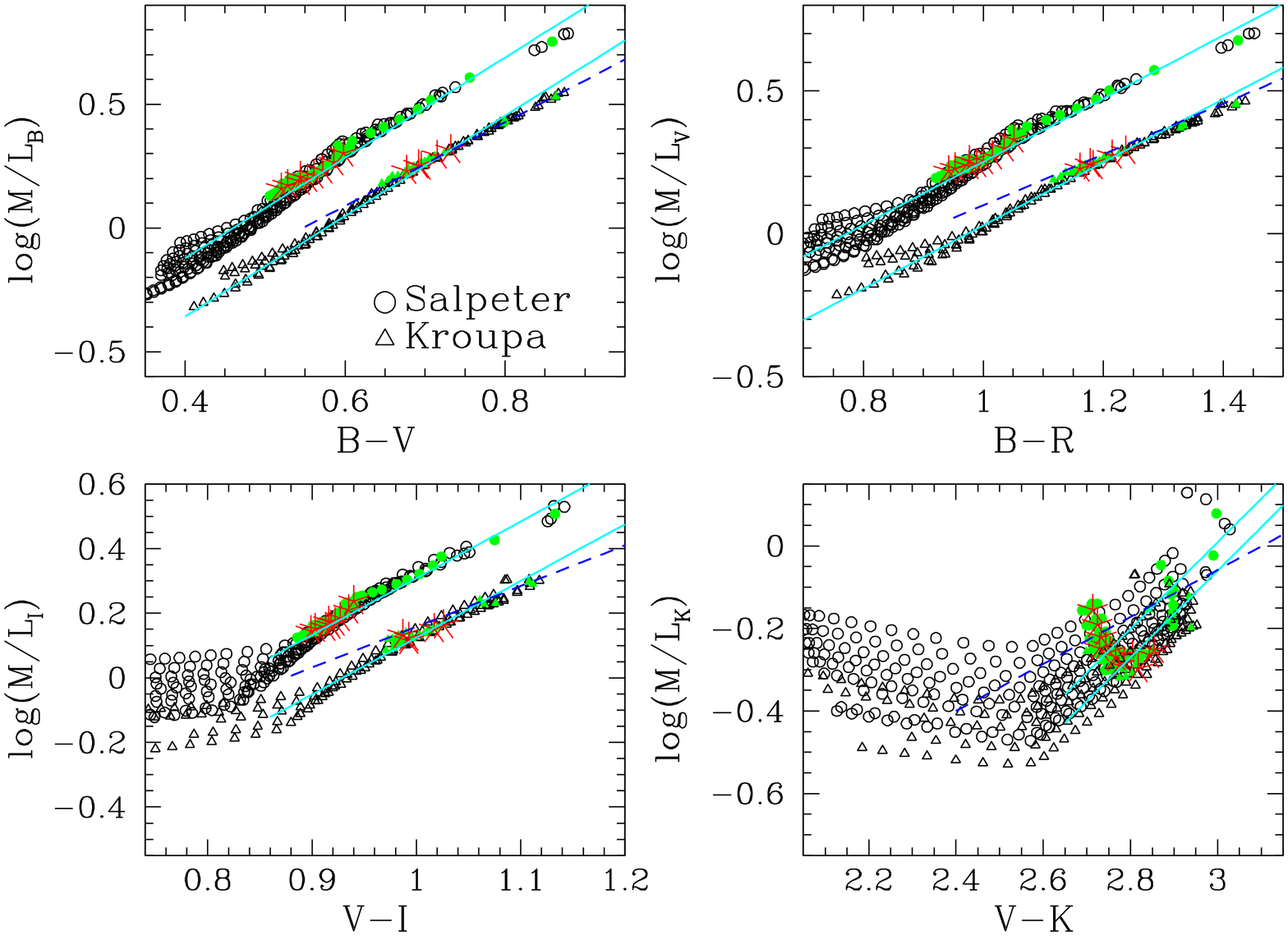}
\caption{CMLR for the chemo--photometric models of disc galaxies of 
PST04 with Salpeter and Kroupa IMF, updated with the new SSP set. 
Symbols as in Fig.~B1 of PST04. Open symbols: one--zone models for individual 
disc annuli; filled symbols: multi--zone models (integrated photometry 
of bulge region, disc region, or global galaxy); asterisks, ``optical disc'' 
region; see PST04 for more details. The solid lines represent the linear fits
for the different IMFs; the dashed line is the superseded CMLR of PST04 for the
Kroupa IMF.}
\label{fig:CMLR_PST04}
\end{center}
\end{figure*}
%%%%%%%%%%%%%%%%%%%%%%%%%%%%%%%%%%%%%%%%%%%%

The ``wedge'' pattern in the $M_\star/L_K$--$(B-R)$ relation, typical of 
population synthesis models of the early 2000's, directly reflected in the 
multi--band analysis of the galaxy mass function of 
\citet{Bell_apjss_149}. In Fig.~\ref{fig:bell03} we reproduce their Fig.~20, 
where each galaxy was assigned a location in the CMLR plane by
$\chi^2$--optimization within the underlying grid of exponential SFH models; 
to this we overlay the theoretical grid 
from Fig.~2 of BdJ, vertically shifted to adjust it to the IMF normalization 
adopted by Bell et~al.\footnote{Although the model grid of
\citet{Bell_apjss_149}
was based on the PEGASE package rather than on GALAXEV, \citet{Maraston_mn_362}
showed that the PEGASE and GALAXEV synthesis models shared the same limits
as to TP-AGB phase implementation.}
Clearly the colour--$M_\star/L$ pattern predicted at the time beared on
the estimated $M_\star/L$ ratios and on the resulting
``semi--empirical'' CMLR of \citet{Bell_apjss_149}, that is much flatter 
than the original theoretical relation of BdJ and has been widely used 
thereafter. Considering the rather different CMLR predicted by modern models, 
and the crucial role of NIR
light in determining stellar mass, those results on the stellar mass function 
and semi--empirical CMLR are worth a revision.

Also, we argue against the blind application of 
semi-empirical relations established for the galaxy population as a whole, 
when interpreting photometric properties within individual galaxies 
\citep[e.g.][]{Kassin_apj_643}. The very flat slope of the semi--empirical 
CMLR of \citet{Bell_apjss_149} is evidently driven by a population of
objects with very blue optical colours but high NIR $M_\star/L$ ratios, 
that stands aside of the rest of the trend as very metal--poor (dwarf?) 
galaxies with old stellar populations. 
Their chemical and 
photometric properties are then quite different from those found within normal 
disc galaxies, so that CMLR derived including these objects do not apply
to colour profiles in spiral galaxies.

%%%%%%%%%%%%%%%%%%%%%%%%%%%%%%%%%%%%%%%%%%%% Figure 11
\begin{figure*}
\begin{center}
\includegraphics[scale=1]{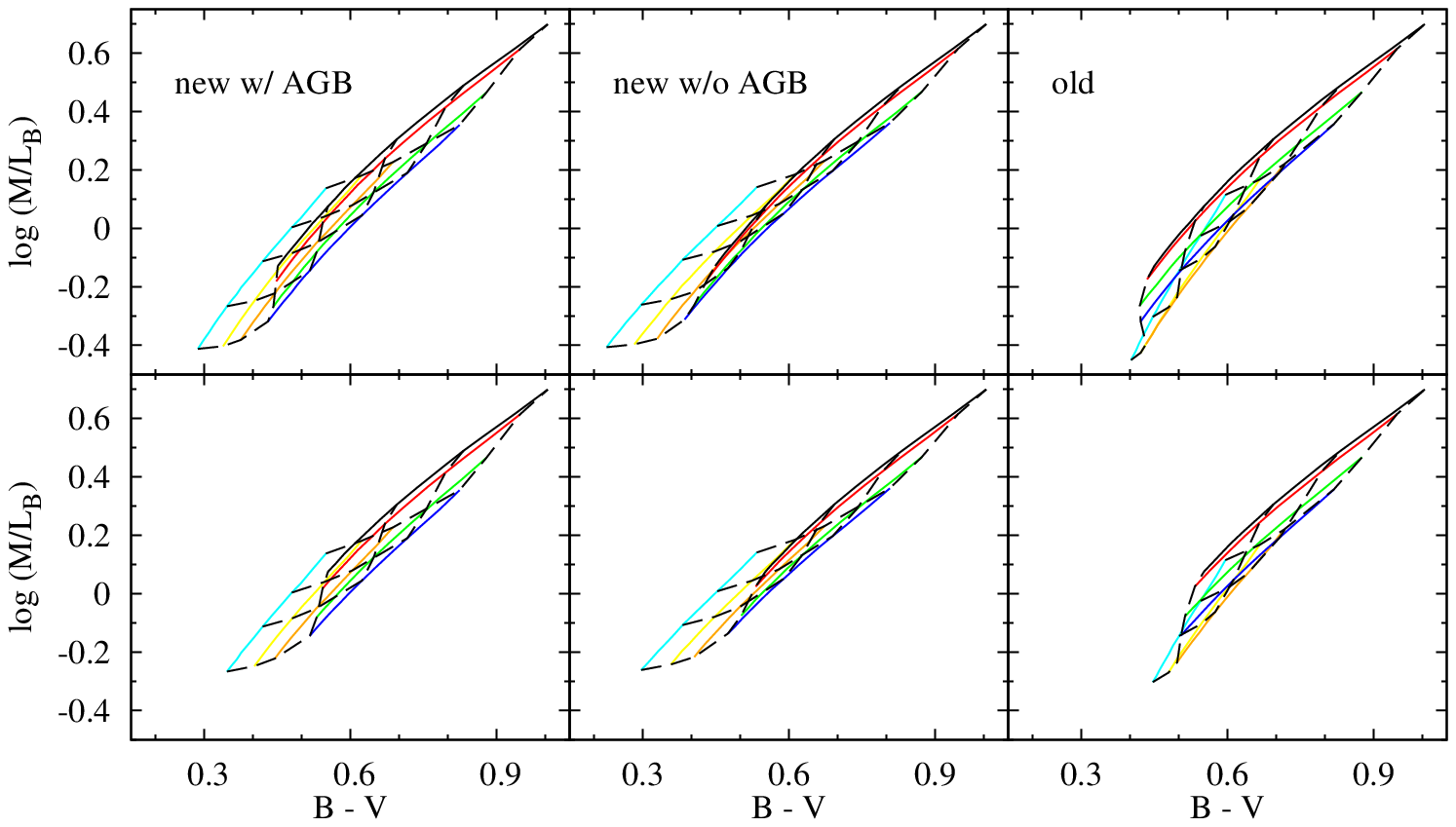}
\caption{CMLR for $M_\star/L_B$ vs.\ $(B-V)$ for the exponential models
with $b \leq 2$ (top panels) or $b \leq 1$ (bottom panels). {\it Left}:
CMLR from the ``new'' SSP dataset; {\it middle}: excluding the contribution
of the TP-AGB phase; {\it right}: CMLR from the ``old'' dataset.}
\label{fig:CMLR_break}
\end{center}
\end{figure*}
%%%%%%%%%%%%%%%%%%%%%%%%%%%%%%%%%%%%%%%%%%%%

%%%%%%%%%%%%%%%%%%%%%%%%%%%%%%%%%%%%%%%%%%%%%%%%%%%%%%%%%%%%%%%%%%%%%%%%%%%%%%%%
\section{Colour--$M_\star/L$ relations for disc galaxy models}
\label{sect:PST04}

In this section we recompute the CMLR for the disc galaxy models of
\citet[][hereinafter PST04]{Portinari_mn_347} and 
discuss the updated CMLR relevant {\it within} individual galaxies.

Fig.~\ref{fig:CMLR_PST04} reproduces Fig.~B1 of PST04 with the new
photometric models --- showing only the Salpeter and Kroupa IMF case for
clarity. The old CMLR of PST04 for the Kroupa IMF case are shown as dashed 
lines for comparison. 
For purely optical CMLR (top panels) the new and old 
CMLR differ at the blue end: the old CMLR were linear down to a certain break 
point in colour (the blue limit of the dashed lines; see also
Fig.~B1 in PST04), below which $M_\star/L$ quickly dropped.
In the new CMLR, the relation steepens toward the blue (say, below 0.6 in 
$B-V$, 0.9 in $B-R$) with no abrupt break; this smoother trend reflects 
that of the new SSPs in Fig.~\ref{fig:ssp_comparison} (top mid panel). 
Still, a single linear fit is adequate over a wide colour range, with a slope 
somewhat steeper than in the CMLR of PST04. The fitting coefficients are listed 
in Table~\ref{tab:fit_PST04} and Table~\ref{tab:fit_PST04_sdss}.
(Notice that disc models cover a smaller colour range than the exponential 
models, as SFHs and metallicities do not get as extreme as considered 
in the previous section.)

The break/steepening of the optical CMLR at the blue end seems supported 
by observations \citep{McGaugh2005} so it is worth commenting in detail.
In the inside--out scenario, disc galaxies are characterized by radial 
gradients in metallicity and SFH: blue colours in the disc outskirts result
from a combination of low metallicities {\it and} slow SFHs ($b \simeq 1$). 
To illustrate the consequence of this folding, 
Fig.~\ref{fig:CMLR_break} displays the $M_\star/L_B - (B-V)$ relation 
from exponential models. For SFHs typical of ``normal'' galaxies ($b \leq 1$, 
bottom panels) around $(B-V)=0.5$ there's a shift from $Z \geq 0.004$ to 
$Z < 0.004$: bluer colours can only be obtained for lower metallicities, 
which for a given SFH (e.g., $b=1$) have systematically lower optical 
$M_\star/L$. Indeed, the old CMLR of PST04 had a break in $M_\star/L$ at 
$(B-V) \simeq 0.5$. The effect is stronger when the
TP-AGB phase is properly included (left panels vs.\ mid panels)  
and was even stronger with the old TP-AGB
prescriptions of \citet{Girardi_mn_300} (right panels), that led to the 
abrupt break in the CMLR of PST04 --- while no such effect was seen 
in the BdJ models for global galaxies.
This shows that, albeit linear $\log M_\star/L$---colour relations are 
a reasonable and handy approximation over a very large range in colours, 
for detailed studies it is worth to consider realistic chemical evolution 
models and the role of the mass--metallicity relation (for galaxies in general)
or of metallicity gradients (within individual objects).

As the drop in $M_\star/L$ at the blue end is favoured by the data 
\citep{McGaugh2005}, it would be interesting to apply the new, smoother 
but steeper CMLR.
In any case, in the outskirts of disc galaxies the $M_\star/L$ ratio should 
be lower than predicted by the BdJ recipe 
\citep[and by the even flatter semi--empirical relations 
of][]{Bell_apjss_149}.

When considering red or NIR bands (bottom panels in 
Fig.~\ref{fig:CMLR_PST04}), the CMLR relation flattens out at the blue end, 
with a large scatter. This is also consequence of the metallicity gradients:
blue colours correspond to the metal--poor outskirts of discs, and 
at low $Z$ red--NIR $M_\star/L$ ratios tend to be larger and do not follow
tight CMLR (Fig.~\ref{fig:exp_BRIK}, left panels). In Table~\ref{tab:fit_PST04}
and~\ref{tab:fit_PST04_sdss} we indicate the blue limit for the log--linear 
CMLR, below which flattening occurs and scatter in $M_\star/L$ at given 
colour becomes significant, exceeding $\pm 0.1$~dex.

The bottom right panel of Fig.~\ref{fig:CMLR_PST04} shows $M_\star/L_K$ 
vs.\ $(V-K)$, representative of optical--NIR CMLR in general.
As $(V-K)$ is mostly a metallicity tracer, rather than a $M_\star/L$ tracer,
the new relation is very steep at the red end
(central regions of disc galaxies with high metallicities) and flattens out
below $(V-K) = 2.6$, presenting everywhere a large scatter.
Though we still provide CMLR based on optical--NIR 
colours in Table~\ref{tab:fit_PST04}, we remark with asterisks that these 
colours are not good tracers of $M_\star/L$.

\begin{figure}
\begin{center}
\includegraphics[scale=0.26,angle=270]{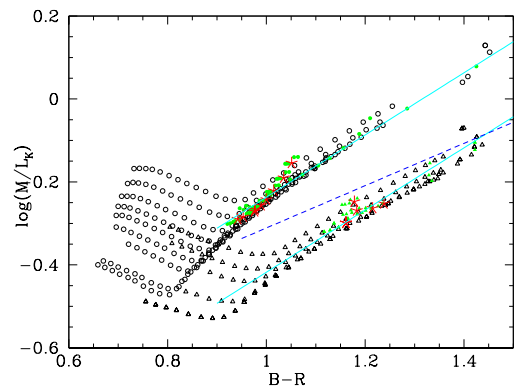}
\caption{$M_\star/L_K$ vs.\ $(B-R)$ relation for the disc galaxy models of PST04.
Lines and symbols as in Fig.~\protect{\ref{fig:CMLR_PST04}}.}
\label{fig:CMLR_BR_MLK}
\end{center}
\end{figure}

Optical colours should always be preferred --- even when estimating NIR 
$M_\star/L$ (Fig.~\ref{fig:CMLR_BR_MLK}). The new
optical colour---NIR $M_\star/L$ relations are steeper,
with lower $M_\star/L$ at a given colour; and 
are nicely tight down to some blue limit ($\sim$0.9 for $B-R$) 
below which they display the typical flattening and large scatter of the low 
metallicity regimes. 

We remark again that, when studying the profiles of disc galaxies (for the sake
of decomposing rotation curves, for instance) this sort of steep CMLR should be
adopted, as they correspond to self--consistent folding of SF and metal 
enrichment histories, with low metallicities always associated to slow 
SFH and young stellar populations in the outer regions. 
``Global'' CMLR for the
galaxy population as a whole \citep{Bell_apjss_149} are much flatter due to 
the contribution of galaxies with metal--poor old populations 
(Fig.~\ref{fig:bell03}), that have no counterpart in disc galaxies. 
Depending on the type of problem at hand, the suitable set of CMLR (``local''
or ``global'', theoretical or semi--empirical) should be adopted.

%%%%%%%%%%%%%%%%%%%%%%%%%%%%%%%%%%%%%%%%%%%% Figure 12
%%%%%%%%%%%%%%%%%%%%%%%%%%%%%%%%%%%%%%%%%%%%

%%%%%%%%%%%%%%%%%%%%%%%%%%%%%%%%%%%%%%%%%% Table 
\begin{table*}
\caption{Fitting coefficients of CMLR of the form : 
$\log (M_\star/L) = s \times {\rm colour} + z$ from disc galaxy models.
Zero--points are given for the Kroupa and the Salpeter IMF normalizations.
Asterisks warn against those colours that are not optimal $M_\star/L$ tracers.
For NIR $M_\star/L$ ratios, we also indicate the blue limit below which the
relation flattens out (see text). The colour range effectively covered by the
disc galaxy models is indicated at the bottom of each sub-table.}
\label{tab:fit_PST04}
\begin{scriptsize}
\begin{tabular}{lccccc}
\hline
\hline
colour  & $M_\star/L$ &  $s$  & $z_{Kr}$ & $z_{Salp}$ & blue lim \\
\hline
$(B-V)$ &        $B$        & 2.027 & -1.168  &  -0.933   &          \\
$(B-V)$ &        $V$        & 1.627 & -0.900  &  -0.665   &          \\
$(B-V)$ &        $R$        & 1.438 & -0.811  &  -0.580   &          \\
$(B-V)$ &        $I$        & 1.294 & -0.782  &  -0.559   &          \\
$(B-V)$ &        $J$        & 1.074 & -0.832  &  -0.634   &   $>$0.5 \\
$(B-V)$ &        $H$        & 1.042 & -0.958  &  -0.770   &   $>$0.5 \\
$(B-V)$ &        $K$        & 1.070 & -1.023  &  -0.836   &   $>$0.5 \\
$(B-V)$ &      $J^{2M}$      & 1.050 & -0.812  &  -0.618   &   $>$0.5 \\
$(B-V)$ &      $H^{2M}$      & 1.032 & -0.953  &  -0.767   &   $>$0.5 \\
$(B-V)$ &      $Ks^{2M}$     & 1.054 & -1.020  &  -0.837   &   $>$0.5 \\
0.35--0.9 & & & & & \\
\hline
$(B-R)$ &        $B$        & 1.374 & -1.384  &  -1.164   &          \\
$(B-R)$ &        $V$        & 1.106 & -1.077  &  -0.855   &          \\
$(B-R)$ &        $R$        & 0.974 & -0.964  &  -0.744   &          \\
$(B-R)$ &        $I$        & 0.870 & -0.912  &  -0.699   &          \\
$(B-R)$ &        $J$        & 0.753 & -0.976  &  -0.785   &   $>$0.9 \\ 
$(B-R)$ &        $H$        & 0.731 & -1.098  &  -0.916   &   $>$0.9 \\ 
$(B-R)$ &        $K$        & 0.750 & -1.168  &  -0.987   &   $>$0.9 \\ 
$(B-R)$ &     $J^{2M}$       & 0.735 & -0.953  &  -0.765   &   $>$0.9 \\
$(B-R)$ &     $H^{2M}$       & 0.723 & -1.093  &  -0.913   &   $>$0.9 \\
$(B-R)$ &     $Ks^{2M}$      & 0.739 & -1.162  &  -0.985   &   $>$0.9 \\
0.7--1.45 & & & & & \\
\hline
$(B-I)$ &        $B$        & 1.141 & -1.688  &  -1.480   &   $>$1.25 \\
$(B-I)$ &        $V$        & 0.916 & -1.317  &  -1.103   &   $>$1.25 \\
$(B-I)$ &        $R$        & 0.814 & -1.187  &  -0.974   &   $>$1.25 \\
$(B-I)$ &        $I$        & 0.741 & -1.136  &  -0.928   &   $>$1.25 \\
$(B-I)$ &        $J$        & 0.638 & -1.166  &  -0.977   &   $>$1.40 \\
$(B-I)$ &        $H$        & 0.627 & -1.297  &  -1.115   &   $>$1.40 \\
$(B-I)$ &        $K$        & 0.641 & -1.367  &  -1.187   &   $>$1.40 \\
$(B-I)$ &     $J^{2M}$       & 0.627 & -1.145  &  -0.958   &   $>$1.40 \\
$(B-I)$ &     $H^{2M}$       & 0.625 & -1.297  &  -1.116   &   $>$1.40 \\
$(B-I)$ &     $Ks^{2M}$      & 0.638 & -1.370  &  -1.192   &   $>$1.40 \\
1.0--2.0 & & & & & \\
\hline
$(B-K)$ &        $B$        & 0.892 & -2.860  &  -2.741   &   $>$3.0  \\
$(B-K)$ &        $V$        & 0.719 & -2.269  &  -2.125   &   $>$3.0  \\
$(B-K)$ &        $R$        & 0.643 & -2.046  &  -1.895   &   $>$3.0  \\
$(B-K)$ &        $I$        & 0.592 & -1.942  &  -1.789   &   $>$3.0  \\
$(B-K)$* &       $J$        & 0.508 & -1.855  &  -1.721   &   $>$3.1  \\
$(B-K)$* &       $H$        & 0.495 & -1.960  &  -1.833   &   $>$3.1  \\
$(B-K)$* &       $K$        & 0.504 & -2.037  &  -1.914   &   $>$3.1  \\
2.4--3.9 & & & & & \\
\hline
$(B-Ks)$ &       $B$        & 0.903 & -2.950  &  -2.831   &   $>$3.1  \\
$(B-Ks)$ &       $V$        & 0.724 & -2.330  &  -2.185   &   $>$3.1  \\
$(B-Ks)$ &       $R$        & 0.646 & -2.096  &  -1.944   &   $>$3.1  \\
$(B-Ks)$ &       $I$        & 0.595 & -1.988  &  -1.834   &   $>$3.1  \\
$(B-Ks)$* &    $J^{2M}$      & 0.509 & -1.886  &  -1.754   &   $>$3.2  \\
$(B-Ks)$* &    $H^{2M}$      & 0.507 & -1.034  &  -0.908   &   $>$3.2  \\
$(B-Ks)$* &    $Ks^{2M}$     & 0.516 & -2.116  &  -1.995   &   $>$3.2  \\
2.5--3.9 & & & & & \\
\hline
$(V-R)$  &       $B$        & 4.321 & -1.867  &  -1.682   &   $>$0.38 \\
$(V-R)$  &       $V$        & 3.478 & -1.465  &  -1.270   &   $>$0.38 \\
$(V-R)$  &       $R$        & 3.078 & -1.313  &  -1.118   &   $>$0.38 \\
$(V-R)$  &       $I$        & 2.773 & -1.235  &  -1.044   &   $>$0.38 \\
$(V-R)$  &       $J$        & 2.430 & -1.270  &  -1.101   &   $>$0.40 \\ 
$(V-R)$  &       $H$        & 2.311 & -1.359  &  -1.201   &   $>$0.40 \\ 
$(V-R)$  &       $K$        & 2.386 & -1.442  &  -1.285   &   $>$0.40 \\ 
$(V-R)$  &    $J^{2M}$       & 2.353 & -1.230  &  -1.064   &   $>$0.40 \\
$(V-R)$  &    $H^{2M}$       & 2.267 & -1.341  &  -1.185   &   $>$0.40 \\ 
$(V-R)$  &    $Ks^{2M}$      & 2.315 & -1.415  &  -1.263   &   $>$0.40 \\
0.32--0.58 & & & & & \\
\hline
$(V-I)$  &       $B$        & 2.677 & -2.423  &  -2.253   &   $>$0.85  \\
$(V-I)$  &       $V$        & 2.156 & -1.916  &  -1.732   &   $>$0.85  \\
$(V-I)$  &       $R$        & 1.922 & -1.724  &  -1.538   &   $>$0.85  \\
$(V-I)$  &       $I$        & 1.756 & -1.632  &  -1.448   &   $>$0.85  \\
$(V-I)$  &       $J$        & 1.549 & -1.629  &  -1.464   &   $>$0.88 \\
$(V-I)$  &       $H$        & 1.508 & -1.738  &  -1.580   &   $>$0.88 \\
$(V-I)$  &       $K$        & 1.546 & -1.822  &  -1.667   &   $>$0.88 \\
$(V-I)$  &    $J^{2M}$       & 1.513 & -1.592  &  -1.429   &   $>$0.88 \\
$(V-I)$  &    $H^{2M}$       & 1.495 & -1.729  &  -1.573   &   $>$0.88 \\
$(V-I)$  &    $Ks^{2M}$      & 1.528 & -1.812  &  -1.659   &   $>$0.88 \\
0.70--1.14 & & & & & \\
\hline
\end{tabular}
\begin{tabular}{lccccc}
\hline
\hline
colour  & $M_\star/L$ &  $s$  & $z_{Kr}$ & $z_{Salp}$ & blue lim \\
\hline
$(V-J)$* &       $B$        & 1.648 &  -2.924 &  -2.862   &  $>$1.7 \\
$(V-J)$* &       $V$        & 1.348 &  -2.358 &  -2.259   &  $>$1.7 \\
$(V-J)$* &       $R$        & 1.213 &  -2.141 &  -2.030   &  $>$1.7 \\
$(V-J)$* &       $I$        & 1.126 &  -2.047 &  -1.930   &  $>$1.7 \\
$(V-J)$* &       $J$        & 0.992 &  -1.991 &  -1.890   &  $>$1.8 \\
$(V-J)$* &       $H$        & 0.985 &  -2.127 &  -2.031   &  $>$1.8 \\
$(V-J)$* &       $K$        & 1.009 &  -2.220 &  -2.128   &  $>$1.8 \\
1.35--2.15 & & & & & \\
\hline
$(V-J^{2M})$* &   $B$        & 1.478 &  -2.669 &  -2.611   &  $>$1.75 \\
$(V-J^{2M})$* &   $V$        & 1.197 &  -2.126 &  -2.031   &  $>$1.75 \\
$(V-J^{2M})$* &   $R$        & 1.072 &  -1.923 &  -1.815   &  $>$1.75 \\
$(V-J^{2M})$* &   $I$        & 0.988 &  -1.832 &  -1.718   &  $>$1.75 \\
$(V-J^{2M})$* & $J^{2M}$      & 0.947 &  -1.951 &  -1.850   &  $>$1.85 \\
$(V-J^{2M})$* & $H^{2M}$      & 0.959 &  -2.129 &  -2.033   &  $>$1.85 \\
$(V-J^{2M})$* & $Ks^{2M}$     & 0.982 &  -2.226 &  -2.135   &  $>$1.85 \\
1.4--2.2 & & & & & \\
\hline
$(V-H)$* &       $B$        & 1.610 &  -4.011 &  -3.969   &  $>$2.45 \\
$(V-H)$* &       $V$        & 1.336 &  -3.297 &  -3.214   &  $>$2.45 \\
$(V-H)$* &       $R$        & 1.180 &  -2.927 &  -2.830   &  $>$2.45 \\
$(V-H)$* &       $I$        & 1.094 &  -2.773 &  -2.670   &  $>$2.45 \\
$(V-H)$* &       $J$        & 0.981 &  -2.679 &  -2.589   &  $>$2.50 \\
$(V-H)$* &       $H$        & 0.979 &  -2.823 &  -2.738   &  $>$2.50 \\
$(V-H)$* &       $K$        & 1.000 &  -2.925 &  -2.844   &  $>$2.50 \\
1.9--2.9 & & & & & \\
\hline
$(V-H^{2M})$* &   $B$        & 1.571 &  -3.956 &  -3.917   &  $>$2.45 \\
$(V-H^{2M})$* &   $V$        & 1.282 &  -3.194 &  -3.114   &  $>$2.45 \\
$(V-H^{2M})$* &   $R$        & 1.153 &  -2.892 &  -2.797   &  $>$2.45 \\
$(V-H^{2M})$* &   $I$        & 1.070 &  -2.744 &  -2.643   &  $>$2.45 \\
$(V-H^{2M})$* & $J^{2M}$      & 0.941 &  -2.599 &  -2.512   &  $>$2.50 \\
$(V-H^{2M})$* & $H^{2M}$      & 0.954 &  -2.789 &  -2.707   &  $>$2.50 \\
$(V-H^{2M})$* & $Ks^{2M}$     & 0.976 &  -2.899 &  -2.822   &  $>$2.50 \\
1.95--2.90 & & & & & \\
\hline
$(V-K)$* &       $B$        & 1.689 &  -4.465 &  -4.443   &  $>$2.60 \\
$(V-K)$* &       $V$        & 1.373 &  -3.595 &  -3.528   &  $>$2.60 \\
$(V-K)$* &       $R$        & 1.231 &  -3.242 &  -3.159   &  $>$2.60 \\
$(V-K)$* &       $I$        & 1.141 &  -3.064 &  -2.973   &  $>$2.60 \\
$(V-K)$* &       $J$        & 1.034 &  -2.970 &  -2.892   &  $>$2.65 \\
$(V-K)$* &       $H$        & 1.035 &  -3.123 &  -3.050   &  $>$2.65 \\
$(V-K)$* &       $K$        & 1.054 &  -3.222 &  -3.153   &  $>$2.65 \\
2.05--3.00 & & & & & \\
\hline
$(V-Ks^{2M})$* &  $B$        & 1.628 &  -4.382 &  -4.363   &  $>$2.65 \\
$(V-Ks^{2M})$* &  $V$        & 1.331 &  -3.549 &  -3.485   &  $>$2.65 \\
$(V-Ks^{2M})$* &  $R$        & 1.198 &  -3.214 &  -3.134   &  $>$2.65 \\
$(V-Ks^{2M})$* &  $I$        & 1.113 &  -3.046 &  -2.958   &  $>$2.65 \\
$(V-Ks^{2M})$* &  $J^{2M}$    & 0.979 &  -2.867 &  -2.791   &  $>$2.72 \\
$(V-Ks^{2M})$* &  $H^{2M}$    & 0.991 &  -3.056 &  -2.985   &  $>$2.72 \\
$(V-Ks^{2M})$* &  $Ks^{2M}$   & 1.015 &  -3.176 &  -3.110   &  $>$2.72 \\
2.1--3.1 & & & & & \\
\hline
$(R-I)$  &       $B$        & 7.003 &  -3.315 &  -3.172   &  $>$0.45 \\
$(R-I)$  &       $V$        & 5.614 &  -2.620 &  -2.457   &  $>$0.45 \\
$(R-I)$  &       $R$        & 5.005 &  -2.354 &  -2.186   &  $>$0.45 \\
$(R-I)$  &       $I$        & 4.605 &  -2.223 &  -2.054   &  $>$0.45 \\
$(R-I)$  &       $J$        & 3.964 &  -2.099 &  -1.955   &  $>$0.47 \\
$(R-I)$  &       $H$        & 3.862 &  -2.196 &  -2.059   &  $>$0.47 \\
$(R-I)$  &       $K$        & 3.955 &  -2.290 &  -2.156   &  $>$0.47 \\
$(R-I)$  &       $J^{2M}$    & 3.874 &  -2.052 &  -1.910   &  $>$0.47 \\
$(R-I)$  &       $H^{2M}$    & 3.834 &  -2.186 &  -2.050   &  $>$0.47 \\
$(R-I)$  &       $Ks^{2M}$   & 3.915 &  -2.278 &  -2.146   &  $>$0.47 \\
0.37--0.57 & & & & & \\
\hline
 & & & & & \\
 & & & & & \\
 & & & & & \\
 & & & & & \\
 & & & & & \\
 & & & & & \\
 & & & & & \\
 & & & & & \\
 & & & & & \\
 & & & & & \\
 & & & & & \\
 & & & & & \\
\end{tabular}
\end{scriptsize}
\end{table*}
%%%%%%%%%%%%%%%%%%%%%%%%%%%%%%%%%%%%%%%%%%

%%%%%%%%%%%%%%%%%%%%%%%%%%%%%%%%%%%%%%%%%% Table 
\begin{table}
\caption{Same as Table~\protect{\ref{tab:fit_PST04}} but for SDSS colours.}
%The colour range effectively covered by the
%disc galaxy models is indicated at the bottom of each sub-table.}
\label{tab:fit_PST04_sdss}
\begin{footnotesize}
\begin{tabular}{lccccc}
\hline
\hline
colour  & $\log (M_\star/L)$ &  $s$  & $z_{Kr}$ & $z_{Salp}$ & blue lim \\
\hline
$(g-r)$ &        $g$        & 1.930 &  -0.851  & -0.634   &          \\
$(g-r)$ &        $r$        & 1.530 &  -0.663  & -0.445   &          \\
$(g-r)$ &        $i$        & 1.370 &  -0.633  & -0.420   &          \\
$(g-r)$ &        $z$        & 1.292 &  -0.665  & -0.462   &   $>$0.3 \\
$(g-r)$ &        $J^{2M}$    & 1.139 &  -0.732  & -0.544   &   $>$0.4 \\
$(g-r)$ &        $H^{2M}$    & 1.128 &  -0.880  & -0.699   &   $>$0.4 \\
$(g-r)$ &        $Ks^{2M}$   & 1.153 &  -0.945  & -0.767   &   $>$0.4 \\
0.25--0.75 & & & & & \\
\hline
$(g-i)$ &        $g$        & 1.385 &  -0.899  & -0.698   &   $>$0.5 \\
$(g-i)$ &        $r$        & 1.098 &  -0.702  & -0.498   &   $>$0.5 \\
$(g-i)$ &        $i$        & 0.985 &  -0.669  & -0.468   &   $>$0.5 \\
$(g-i)$ &        $z$        & 0.898 &  -0.675  & -0.484   &   $>$0.5 \\
$(g-i)$ &        $J^{2M}$    & 0.868 &  -0.804  & -0.621   &   $>$0.6 \\
$(g-i)$ &        $H^{2M}$    & 0.861 &  -0.952  & -0.776   &   $>$0.6 \\ 
$(g-i)$ &        $Ks^{2M}$   & 0.879 &  -1.019  & -0.847   &   $>$0.6 \\
0.35--1.05 & & & & & \\
\hline
$(g-z)$ &        $g$        & 1.294 &  -1.097  & -0.908   &   $>$0.7 \\ 
$(g-z)$ &        $r$        & 1.033 &  -0.867  & -0.669   &   $>$0.7 \\ 
$(g-z)$ &        $i$        & 0.946 &  -0.838  & -0.642   &   $>$0.7 \\ 
$(g-z)$ &        $z$        & 0.894 &  -0.861  & -0.672   &   $>$0.7 \\ 
$(g-z)$ &        $J^{2M}$    & 0.789 &  -0.906  & -0.734   &   $>$0.8 \\
$(g-z)$ &        $H^{2M}$    & 0.787 &  -1.060  & -0.893   &   $>$0.8 \\ 
$(g-z)$ &        $Ks^{2M}$   & 0.803 &  -1.128  & -0.965   &   $>$0.8 \\ 
0.7--1.3 & & & & & \\
\hline
$(r-i)$ &        $g$        & 5.926 &  -1.293  & -1.116   &   $>$0.20 \\
$(r-i)$ &        $r$        & 4.738 &  -1.025  & -0.838   &   $>$0.20 \\
$(r-i)$ &        $i$        & 4.338 &  -0.982  & -0.795   &   $>$0.20 \\
$(r-i)$ &        $z$        & 4.093 &  -0.996  & -0.816   &   $>$0.20 \\
$(r-i)$ &        $J^{2M}$    & 3.684 &  -1.043  & -0.878   &   $>$0.21 \\
$(r-i)$ &        $H^{2M}$    & 3.659 &  -1.192  & -1.033   &   $>$0.21 \\
$(r-i)$ &        $Ks^{2M}$   & 3.740 &  -1.264  & -1.109   &   $>$0.21 \\
0.12--0.32 & & & & & \\
\hline
$(r-z)$ &        $g$        & 3.902 &  -1.592  & -1.466   &   $>$0.38 \\
$(r-z)$ &        $r$        & 3.116 &  -1.263  & -1.116   &   $>$0.38 \\
$(r-z)$ &        $i$        & 2.863 &  -1.205  & -1.054   &   $>$0.38 \\
$(r-z)$ &        $z$        & 2.716 &  -1.214  & -1.067   &   $>$0.38 \\
$(r-z)$ &        $J^{2M}$    & 2.355 &  -1.196  & -1.067   &   $>$0.40 \\
$(r-z)$ &        $H^{2M}$    & 2.344 &  -1.346  & -1.223   &   $>$0.40 \\
$(r-z)$ &        $Ks^{2M}$   & 2.384 &  -1.416  & -1.297   &   $>$0.40 \\
0.22--0.56 & & & & & \\
\hline
\end{tabular}
\end{footnotesize}
\end{table}
%%%%%%%%%%%%%%%%%%%%%%%%%%%%%%%%%%%%%%%%%%

%%%%%%%%%%%%%%%%%%%%%%%%%%%%%%%%%%%%%%%%%%%%%%%%%%%%%%%%%%%%%%%%%%%%%%%%%%%%%%%%
\section{Attenuation by interstellar dust}\label{sec:attenuation}

We have discussed in Section~\ref{sec:circumstellar_dust} that circumstellar 
dust around AGB stars has a negligible effect on optical and NIR CMLR, so that
we can disregard it. In this section we address the effect of interstellar
dust, that both reddens and dims stellar luminosity. 
\citet{Bell_apj_550} argue that, to first order, in optical CMLR the two 
effects compensate each other, as the dust vector runs almost parallel to the 
dust--free CMLR (age--metallicity--dust degeneracy).

We revisit the effects of interstellar dust on optical and NIR
CMLR over galactic scales, taking advantage of recent recipes based on 
detailed radiative transfer models 
\citep{Tuffs_aa_419}. Our goal is to provide CMLR that are statistically
applicable to large galaxy samples that include a range of morphologies,
intrinsic colours and random inclinations.

%%%%%%%%%%%%%%%%%%%%%%%%%%%%%%%%%%%%%%%%%%%%%%%%%%%%%%%%%%%%%%%%%%%%%%%%%%%%%%%
\subsection{The attenuated spiral galaxy models}

We construct simple models of spiral galaxies consisting of bulge+disc, 
and then apply the dust corrections of 
\citet{Tuffs_aa_419}.
The models include a bulge component, represented by a
10~Gyr old SSP, and a disc component with exponential SFH with $b \leq 2$
(Section~\ref{sec:simple_exp_models}):
$b \leq 1$ values represent ``normal'' disc galaxy morphologies, 
and we extend to somewhat bluer objects with recent intensive SF.
For simplicity, the dust--free models were calculated only for solar 
metallicity: we shall see that dust effects on galactic scales
are statistically more relevant than metallicity effects.

The models have a random distribution in $b=0.01 \div 2$ and in
bulge--to--total ratio B/T=$0 \div 0.6$, typical of disc galaxies 
\citep{Allen_mn_371}; the corresponding luminosity ratios (B/T)$_\lambda$ 
entering Eq.~\ref{eq:optical_attenuation} below, are computed
self--consistently from the SFHs for bulge and disc.
Dust attenuation is then added for random inclinations from face-on to 
edge--on, $0 \leq \cos(i) \leq 1$, for a total of 20000 models.
 
Dust attenuation in the various bands is computed
following the prescriptions of \citet{Tuffs_aa_419}, to which the reader 
is referred for all details. Their dust models have been
successfully tested both on multi--band emission of individual galaxies 
\citep{Popescu_aa_362} and on large galaxy surveys \citep{Driver_mn_379}.
The attenuation prescriptions of Tuffs et~al. depend only on the 
properties of interstellar dust, not on the incident stellar radiation field,
and are presented as applicable to real disc galaxies irrespectively of 
their detailed stellar SED; likewise, they are applicable to model disc 
galaxies irrespectively of the specific population synthesis model used 
to generate them.

%%%%%%%%%%%%%%%%%%%%%%%%%%%%%%%%%%%%%%%%%%%% Figure 13
\begin{figure*}
\includegraphics[scale=0.25,angle=-90]{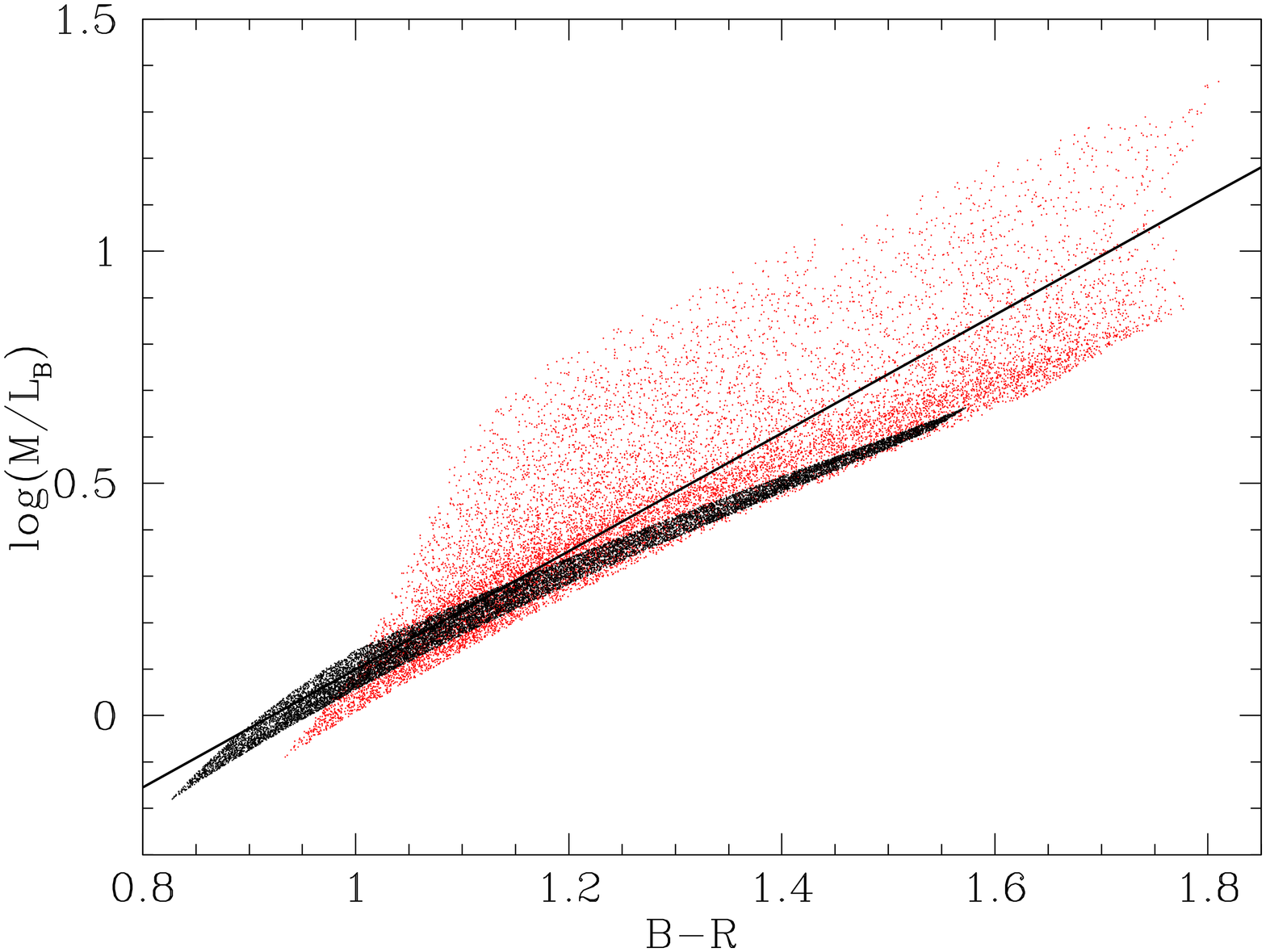}
\includegraphics[scale=0.25,angle=-90]{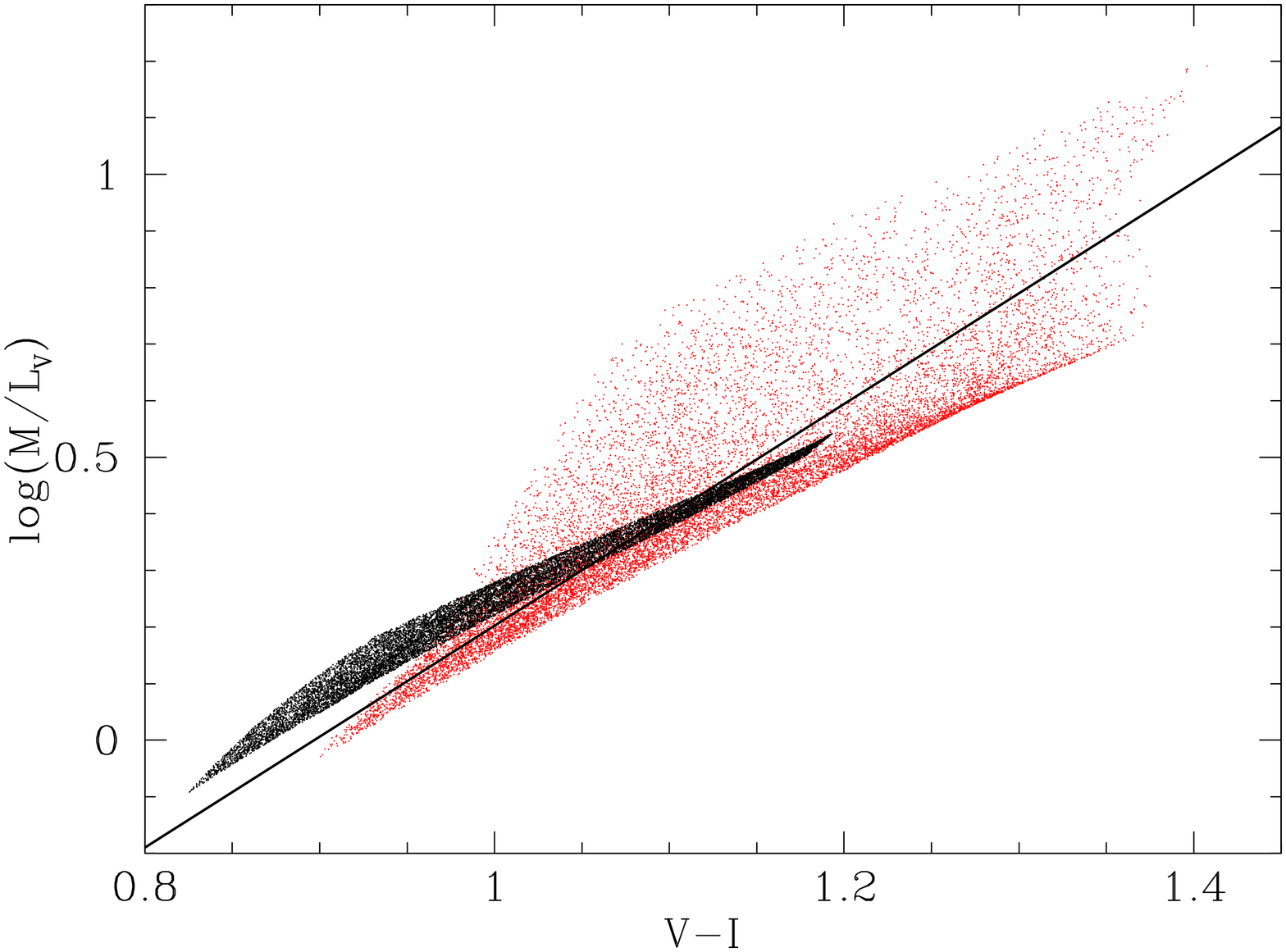}
\includegraphics[scale=0.25,angle=-90]{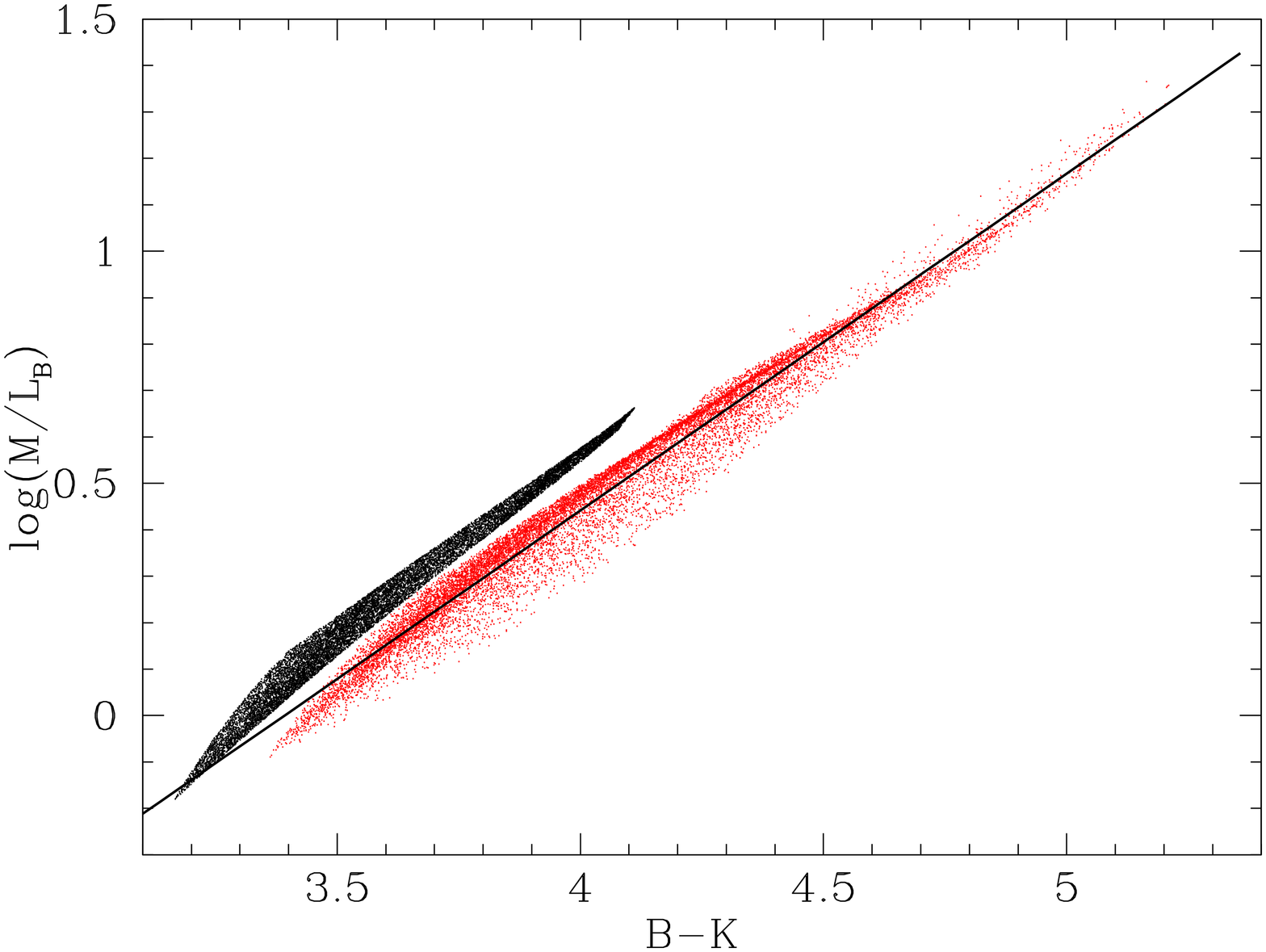}
\includegraphics[scale=0.25,angle=-90]{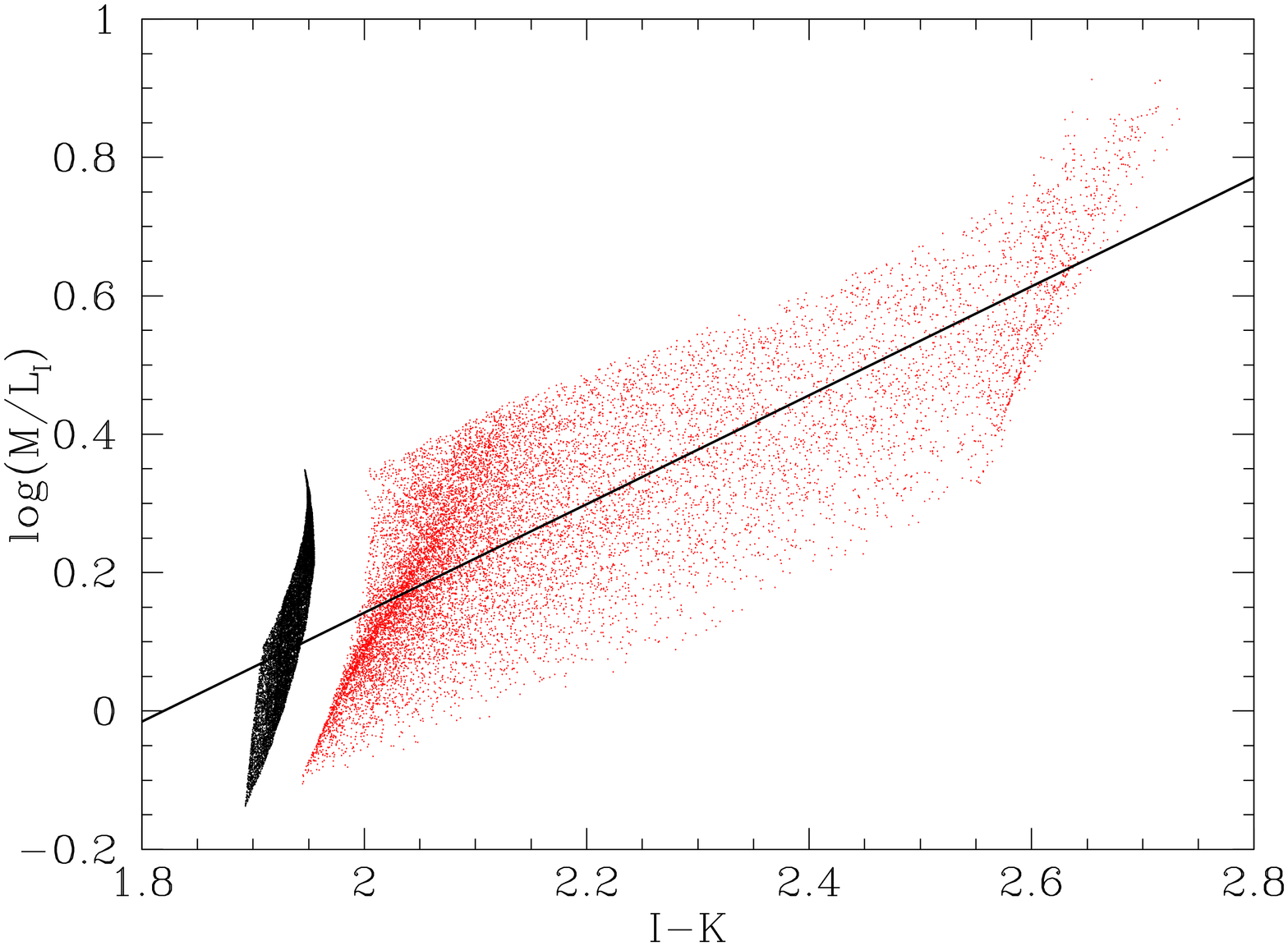}
\caption{CMLR for dust--free model galaxies
(black dots) and attenuated models (red dots). The solid line indicates 
the linear fit to the attenuated models.}
\label{fig:dust}
\end{figure*}
%%%%%%%%%%%%%%%%%%%%%%%%%%%%%%%%%%%%%%%%%%%%

In short, the dust model includes a clumpy dust component associated 
with the star-forming regions in a thin disc \citep[which affects only 
UV light, and can be neglected in the 
optical/NIR;][]{Popescu_aa_362,Tuffs_aa_419} and a diffuse dust component
residing both in the young thin disc and in the older disc. The bulge 
is intrinsically dust--free, but is heavily affected by the dust in the disc, 
to the extent that it turns out to be the most attenuated galaxy component; 
a counter--intuitive result confirmed by observational trends 
\citep{Driver_mn_379}.
The equation for the composite attenuation of the galaxy at wavelength 
$\lambda$ is:
\begin{equation} \label{eq:optical_attenuation}
\Delta m_\lambda = -2.5 \log \Bigg{(} 
r^\mathrm{0,disc}_\lambda 10^{-0.4\Delta m^\mathrm{disc}_\lambda} + 
r^\mathrm{0,bulge}_\lambda 10^{-0.4\Delta m^\mathrm{bulge}_\lambda } \Bigg{)}
\end{equation}
where $r^0_\lambda$ is the fraction of the bulge or disc to the total intrinsic 
luminosity in the band $\lambda$: (B/T)$_\lambda$ and
(1-B/T)$_\lambda$, respectively. (We recast the equation in terms of the 
intrinsic B/T ratio of the galaxy models,
while the original Eq.~18 in Tuffs et~al.\ was expressed in terms of
the apparent, dust--attenuated B/T ratio, to be applied to observed galaxies.)
$\Delta m^\mathrm{disc,bulge}_\lambda$ is the attenuation value in magnitudes 
for the disc or bulge, provided by Tuffs et~al.\ in the 
form of polynomial equations in $(1-\cos i)$. The polynomial coefficients 
are tabulated for various optical and NIR photometric bands, and optical depth 
values $\tau_B^f$.
We use the fixed value $\tau^f_B=4.0$, consistent with the 
empirical mean opacity $\tau_B^f = 3.8 \,\pm \, 0.7$ found by 
\citet{Driver_mn_379} based on 10095 galaxies with bulge-disc decomposition.

We notice in passing that 
Eq.~\ref{eq:optical_attenuation} becomes inconsistent for high B/T values, 
as it predicts an ever--increasing attenuation for increasing bulge fraction,
while in the limit {\mbox{B/T $\longrightarrow$ 1}} attenuation should 
vanish together 
with the disc component. We suspect that this is a consequence of extending 
the disc+dust to the very centre, rather than modelling a hole in the dusty
disc in correspondence to the bulge component. However, here we only consider 
galaxies with an important disc component (B/T $\leq$ 0.6) where the dust 
prescriptions of Tuffs et al.
are well supported by observational evidence \citep{Driver_mn_379}.

%%%%%%%%%%%%%%%%%%%%%%%%%%%%%%%%%%%%%%%%%%%%%%%%%%%%%%%%%%%%%%%%%%%%%%%%%%%%%%%%
\subsection{Colour--$M_\star/L$ relations for dusty galaxies}

Fig.~\ref{fig:dust} shows the effects of dust on various CMLR. 
The dust--free galaxies (black dots) are intrinsically redder with decreasing 
disc $b$--parameter and increasing B/T ratio. They follow, within the quoted 
uncertainty of 0.1~dex, the CMLR predicted by the exponential models of 
Section~\ref{sec:simple_exp_models}: the 
superposition of two components, disc+bulge (exponential SFH + old SSP) 
still closely follow the one--component CMLR ---
but there is hardly any CMLR in $(I-K)$, as discussed in 
Section~\ref{sec:simple_exp_models}.
When dust attenuation is taken into account (red dots), the models spread on
a much wider area in the plot, the brighter (fainter) bound 
corresponding to face--on (edge--on) inclinations. 

The top panels show optical CMLR. The spread in $\log M_\star/L$ at a fixed 
colour is about 0.5~dex (a factor--of--3), with dusty galaxies being 
tipically less luminous than the dust--free case. Notice that the red end 
of the dust--free
distribution is scattered much further to the red by attenuation, than the 
blue end; this is another counter--intuitive result of the bulge component
suffering more extinction than the disc.
The largest departures from the dust--free case correspond to the 
highest inclinations, that are statistically less frequent; altogether the 
{\it statistical} CMLR of attenuated galaxies (solid line) is not too far 
from the dust--free exponential case of Section~\ref{sec:simple_exp_models}.
In the $\log M_\star/L$ vs.\ $(B-R)$ plane the dusty CMLR
is mildly offset (0.1~dex) from the dust--free one, 
toward heavier $M_\star/L_B$ at fixed colour.

The (statistical) effect of dust is smallest for $M_\star/L_V$ vs.\ $(V-I)$:
the attenuated CMLR just shows a bit steeper slope than the 
dust--free one, and the difference is even less than 0.1~dex. This is 
interesting as 
\citet{Gallazzi_Bell2009} have recently selected $(g-i)$ as the best
suited colour for stellar mass estimates in the dust--free case. As this colour
is quite close to $(V-I)$, we argue that dust corrections are small
for CMLR involving {\mbox{$(g-i)$}, so that this is a robust optical 
$M_\star/L$ indicator also with respect to dust effects. 
(We cannot address
the role of dust attenuation directly in SDSS bands, as the Tuffs et~al.\ 
prescriptions are only provided for Johnson bands.)

%%%%%%%%%%%%%%%%%%%%%%%%%%%%%%%%%%%%%%%%%% Table 6
\begin{table*}
\caption{Fitting coefficients of CMLR of the form : 
$\log (M_\star/L) = s \times {\rm colour} + z$ for dusty galaxy models.}
\label{tab:fit_dustymodels}
\begin{tabular}{lccc}
\hline
\hline
colour & $\log (M_\star/L)$ & $s$ & $z$ \\  
\hline
$(B-V)$ & $B$ & 1.836 & -0.905 \\
$(B-V)$ & $V$ & 1.493 & -0.681 \\
$(B-V)$ & $R$ & 1.337 & -0.627 \\
$(B-V)$ & $I$ & 1.228 & -0.646 \\
$(B-V)$ & $J$ & 1.048 & -0.803 \\
$(B-V)$ & $H$ & 0.941 & -0.909 \\
$(B-V)$ & $K$ & 0.866 & -0.926 \\
0.5--1.1 & & & \\
\hline
$(B-R)$ & $B$ & 1.272 & -1.173  \\
$(B-R)$ & $V$ & 1.040 & -0.905  \\
$(B-R)$ & $R$ & 0.934 & -0.832  \\
$(B-R)$ & $I$ & 0.860 & -0.837  \\
$(B-R)$ & $J$ & 0.730 & -0.960  \\
$(B-R)$ & $H$ & 0.652 & -1.047  \\
$(B-R)$ & $K$ & 0.599 & -1.051  \\
0.9--1.8 & & & \\
\hline
$(B-I)$ & $B$ & 1.049 & -1.492 \\
$(B-I)$ & $V$ & 0.859 & -1.167 \\
$(B-I)$ & $R$ & 0.772 & -1.069 \\
$(B-I)$ & $I$ & 0.711 & -1.057 \\
$(B-I)$ & $J$ & 0.602 & -1.142 \\
$(B-I)$ & $H$ & 0.537 & -1.209 \\
$(B-I)$ & $K$ & 0.493 & -1.200 \\
1.4--2.5 & & & \\
\hline
$(B-J)$ & $B$ & 0.886 & -2.203 \\
$(B-J)$ & $V$ & 0.724 & -1.746 \\
$(B-J)$ & $R$ & 0.649 & -1.585 \\
$(B-J)$ & $I$ & 0.597 & -1.527 \\
$(B-J)$ & $J$ & 0.505 & -1.541 \\
$(B-J)$ & $H$ & 0.452 & -1.567 \\
$(B-J)$ & $K$ & 0.416 & -1.532 \\
2.4--3.9 & & & \\
\hline
$(B-H)$ & $B$ & 0.803 & -2.604 \\
$(B-H)$ & $V$ & 0.656 & -2.073 \\
$(B-H)$ & $R$ & 0.588 & -1.876 \\
$(B-H)$ & $I$ & 0.540 & -1.793 \\
$(B-H)$ & $J$ & 0.457 & -1.767 \\
$(B-H)$ & $H$ & 0.409 & -1.771 \\
$(B-H)$ & $K$ & 0.377 & -1.720 \\
3.2--4.8 & & & \\
\hline
$(B-K)$ & $B$ & 0.734 & -2.496 \\
$(B-K)$ & $V$ & 0.600 & -1.984 \\
$(B-K)$ & $R$ & 0.537 & -1.797 \\
$(B-K)$ & $I$ & 0.493 & -1.719 \\
$(B-K)$ & $J$ & 0.418 & -1.705 \\
$(B-K)$ & $H$ & 0.374 & -1.716 \\
$(B-K)$ & $K$ & 0.345 & -1.670 \\
3.4--5.2 & & & \\
\hline
\end{tabular}
\begin{tabular}{lccc}
\hline
\hline
colour & $\log (M_\star/L)$ & $s$ & $z$ \\  
\hline
$(V-R)$ & $B$ & 3.966 & -1.697 \\
$(V-R)$ & $V$ & 3.308 & -1.356 \\
$(V-R)$ & $R$ & 2.953 & -1.227 \\
$(V-R)$ & $I$ & 2.700 & -1.191 \\
$(V-R)$ & $J$ & 2.301 & -1.266 \\
$(V-R)$ & $H$ & 2.071 & -1.328 \\
$(V-R)$ & $K$ & 1.911 & -1.314 \\
0.4--0.7 & & & \\
\hline
$(V-I)$ & $B$ & 2.404 & -2.223 \\
$(V-I)$ & $V$ & 1.959 & -1.756 \\
$(V-I)$ & $R$ & 1.752 & -1.589 \\
$(V-I)$ & $I$ & 1.606 & -1.526 \\
$(V-I)$ & $J$ & 1.365 & -1.547 \\
$(V-I)$ & $H$ & 1.225 & -1.577 \\
$(V-I)$ & $K$ & 1.129 & -1.542 \\
0.9--1.4 & & & \\
\hline
$(V-J)$ & $B$ & 1.587 & -3.129 \\
$(V-J)$ & $V$ & 1.296 & -2.501 \\
$(V-J)$ & $R$ & 1.161 & -2.259 \\
$(V-J)$ & $I$ & 1.066 & -2.143 \\
$(V-J)$ & $J$ & 0.902 & -2.065 \\
$(V-J)$ & $H$ & 0.808 & -2.038 \\
$(V-J)$ & $K$ & 0.744 & -1.966 \\
1.9--2.7 & & & \\
\hline
$(V-H)$ & $B$ & 1.266 & -3.428 \\
$(V-H)$ & $V$ & 1.034 & -2.746 \\
$(V-H)$ & $R$ & 0.927 & -2.479 \\
$(V-H)$ & $I$ & 0.851 & -2.345 \\
$(V-H)$ & $J$ & 0.729 & -2.235 \\
$(V-H)$ & $H$ & 0.645 & -2.191 \\
$(V-H)$ & $K$ & 0.595 & -2.110 \\
2.7--3.8 & & & \\
\hline
$(V-K)$ & $B$ & 1.048 & -2.979 \\
$(V-K)$ & $V$ & 0.856 & -2.379 \\
$(V-K)$ & $R$ & 0.767 & -2.150 \\
$(V-K)$ & $I$ & 0.704 & -2.044 \\
$(V-K)$ & $J$ & 0.597 & -1.981 \\
$(V-K)$ & $H$ & 0.535 & -1.966 \\
$(V-K)$ & $K$ & 0.414 & -1.640 \\
2.9--4.2 & & & \\
\hline
$(I-K)$ & $B$ & 0.970 & -1.632 \\
$(I-K)$ & $V$ & 0.885 & -1.480 \\
$(I-K)$ & $R$ & 0.831 & -1.428 \\
$(I-K)$ & $I$ & 0.786 & -1.430 \\
$(I-K)$ & $J$ & 0.627 & -1.376 \\
$(I-K)$ & $H$ & 0.505 & -1.298 \\
$(I-K)$ & $K$ & 0.386 & -1.114 \\
2.0--2.7 & & & \\
\hline
\end{tabular}
\end{table*}
%%%%%%%%%%%%%%%%%%%%%%%%%%%%%%%%%%%%%%%%%%

The role of dust is remarkable in optical--NIR CMLR
(bottom panels): inclination strongly influences the colours, and the strong
reddening renders the dusty optical--NIR CMLR {\it lighter}, at a given colour,
than the dust--free case.

Since the dust--free optical--NIR CMLR depend on metallicity, 
we checked that changing the metallicity 
of the galaxy models (within a factor of 2--3 from solar, as relevant for 
global galaxy metallicities) has very little impact on the resulting dusty 
CMLR: the effect of dust is dominant over metallicity effects.

In some optical--NIR CMLR, attenuation and reddening combine so as to maintain 
tight CMLR --- albeit different from the dust--free case. The best 
example is $(B-K)$, with a $M_\star/L$ scatter of at most $\pm$0.15~dex in all 
bands (bottom left panel). Also $(V-H)$ is a good $M_\star/L$ tracer for 
dusty galaxies, with a scatter of $\pm$0.15~dex in $M_\star/L_V$ or $M_\star/L_H$
at given $(V-H)$.
Even for the $(I-K) - M_\star/L_I$ relation, the scatter ($\pm$0.2~dex) is no 
worse than that for optical dusty CMLR (bottom right).

In summary, when dust attenuated galaxies are considered, optical--NIR CMLR
are no worse (in terms of scatter) than optical--optical CMLR; in some cases
--- notably $(B-K)$ --- they are even favoured. The best option in dusty
optical--NIR CMLR is to derive $M_\star/L$ in one of the bands involved
in the base colour; a very common case in practice.

%%%%%%%%%%%%%%%%%%%%%%%%%%%%%%%%%%%%%%%%%%%%%%%%%%%%%%%%%%%%%%%%%%%%%%%%%%%%%%%%
\section{Summary and conclusions}
In this paper we rediscuss theoretical colour--stellar mass-to-light relations
(CMLR) in the light of modern populations synthesis models including an accurate
implementation of the TP-AGB phase, and of the effects of interstellar dust
as predicted from radiative transfer models of disc galaxies.

The importance of the AGB phase for the integrated NIR luminosity of
stellar populations has been extensively discussed in recent years 
\citep[][and other papers of the same series]{Maraston_mn_362, Maraston2006, 
Tonini2009}.
In intermediate--age stellar populations (0.3--2 Gyr), TP-AGB stars dominate
the NIR luminosity, lowering the $M_\star/L$ ratio by up to a factor 
of 2 and driving very red optical--NIR colours ($V-K \geq 3$); 
NIR $M_\star/L$ ratios are quite independent of metallicity. Optical 
luminosities and colours, on the other hand, are not severely affected.

We update theoretical CMLR for composite stellar populations by means 
of the latest Padova isochrones (Marigo et~al.\ 2008; Girardi et ~al.\ 2010),
which include a far more refined treatment of the TP-AGB phase than 
the previous dataset \citep{Girardi_mn_300, Girardi_aa_391}.

For most of their evolution after the onset of the AGB at $\sim$100~Myr,
the updated Single Stellar Populations are significanly brighter in the NIR 
(up to 0.5~mag), with correspondingly ``lighter'' $M_\star/L$.
The effect of circumstellar dust on the integrated optical and NIR luminosity
is negligible.

Considering both Simple and composite Stellar Populations (the latter with
exponentially decreasing/increasing star formation histories mimicking the
Hubble sequence, as well as from full chemo--photometric galaxy models) 
we highlight the following characteristics of updated CMLR.
\begin{itemize}
\item
Optical CMLR 
are little affected by the upgrade in the AGB models, and remain tight thanks 
to a strong age--metallicity degeneracy 
\citep[cf.][]{Bell_apj_550}.
\item
The integrated NIR luminosity is increased with minor effects in the optical: 
the resulting NIR $M_\star/L$ is about 0.1~dex lighter, at a given optical 
colour.
\item
The integrated NIR luminosity is also less sensitive to metallicity --- at 
least for $Z \geq 0.004$ which is representative of the bulk of the stellar 
populations in galaxies. This favours a more robust estimate of stellar mass 
from NIR light, with tighter NIR $M_\star/L$---optical colour relations 
(like $M_\star/L_K$ vs.\ $B-R$) compared to previous predictions.
\item
In optical--NIR colours, such as $(V-K)$, the new models are both lighter 
and redder, so that the new CMLR are much lighter (up to 0.3~dex, a factor 
of 2) at a given colour.
\item
As noticed by \citet{Bell_apj_550}, optical--NIR colours like $(V-K)$ or 
$(I-K)$ are mostly metallicity indicators, while being very poor $M_\star/L$ 
tracers; this is even more true with the new models.
\item
The new models suggests a revision of results obtained from multi--band 
analysis of the galaxy population spanning from optical to NIR, including 
semi--empirical CMLR \citep{Bell_apjss_149}.
\item
We argue against the use of semi--empirical CMLR established for the
general galaxy population, when studying individual galaxies: the CMLR 
resulting from a coherent star formation and chemical evolution history 
(and their radial gradients) within a single galaxy is potentially 
different from the CMLR obtained as a statistical average of galaxies.
This is evident for instance in the ``break'' of the CMLR
at blue optical colours, seen in chemo--photometric models of disc galaxies
\citep{Portinari_mn_347} and not when considering a wider range of 
(uncorrelated) SFH and metallicities describing the galaxy population 
in general.
\end{itemize}

We finally warn that recent
observational results suggest that the newest population synthesis 
models may actually overestimate the luminosity contribution of AGB stars 
\citep[Kriek et~al.\ 2010;][]{Zibetti2012}, \nocite{Kriek2010} 
partly due to an excess of rare, luminous AGB stars in the models 
(Melbourne et~al.\ 2010) \nocite{Melbourne2012} and partly due to the effects 
of circumstellar dust (Meidt et~al.\ 2012) \nocite{Meidt2012}.
Future, better calibrated AGB models may converge on CMLR 
intermediate between the ``classic'' ones of the early 2000's and those 
presented here.

All of the above refers to dust--free CMLR. It is usually 
assumed that, at least for optical CMLR, dust is a second--order effect thanks
to the age--metallicity--dust degeneracy \citep{Bell_apj_550}. We revisited
this issue considering a more realistic implementation of dust effects on
galactic scales, and found that dust has a non--negligible role.
\begin{itemize}
\item
The combined effect of reddening and attenuation introduces an enormous scatter
even in optical CMLR: highly inclined disc 
galaxies can be 0.5~dex ``heavier'' at a given colour, than predicted by 
dust--free CMLR. So, for individual galaxies, the CMLR does not 
apply unless good inclination information and dust corrections are available.
\item
Nontheless, we can still define ``dusty'' CMLR that {\it statistically}
apply to large
galaxy samples where we lack detailed morphological and inclination information
for individual objects (but are at least able to distinguish a disc--like
galaxy from a pure spheroid with no dust). In the optical, statistical dusty
CMLR are somewhat heavier (about 0.1~dex) at a given colour, than the dust-free
case.
\item
The smallest change with respect to the dust--free case is found 
for $M_\star/L_V$ vs.\ $(V-I)$. 
This suggests that the $(g-i)$ colour, recently selected as optimal
stellar mass tracer in the dust--free case \citep{Gallazzi_Bell2009}, 
remains a good tracer also when considering (or neglecting!) the effects
of dust.
\item
Dust reddening strongly alters optical--NIR CMLR, making them lighter, 
at given colour, than dust--free CMLR.
\item
For some optical--NIR CMLR, extinction and reddening combine to yield rather
tight dusty CMLR --- albeit very different from the dust--free ones.
$(B-K)$ is an excellent stellar mass tracer for dusty galaxies,
with a scatter in $M_\star/L$ within $\pm$0.15~dex in all bands. $M_\star/L_{V,H}$
vs.\ $(V-H)$ CMLR are also similarly tight. For dusty galaxies, these 
CMLR are better mass tracers than optical--optical CMLR.
\end{itemize}

In Tables~\ref{tab:fit_bmodels} through~\ref{tab:fit_dustymodels} we give 
our updated CMLR for the various cases: exponential models (for those colours 
where defining a CMLR is meaningful, at least for $Z \geq 0.004$); 
chemo--photometric
disc galaxy models, where CMLR result from a consistent convolution of
SF and chemical evolution history, to be best applied within individual 
(dust--free or dust--corrected) galaxies; and ``dusty'' CMLR that can be 
statistically applied to large galaxy samples.

CMLR are a robust, handy tool to estimate 
stellar masses; we can optimize their use by choosing the colour and 
relation most suitable for each specific problem.

\section*{Acknowledgments}
We thank L\'eo Girardi and Paola Marigo for useful discussions on the
Padova stellar models.
This study was financed by the Academy of Finland (grants nr.~130951 and 218317)
and by the Magnus Ehrnrooth foundation.

\bibliographystyle{mn2e}
\bibliography{mnemonics,astrobib}
%\bibliography{astrobib}

\end{document}